\documentclass[a4paper,fleqn,usenatbib]{mnras}

\usepackage{newtxtext,newtxmath}
\usepackage[T1]{fontenc}
\usepackage{ae,aecompl}

\usepackage{graphicx}	                                    % Including figure files
\usepackage{amsmath}	                                    % Advanced maths commands
\usepackage{mathtools}	                                    % To split equation over two lines
\usepackage{array}	                                            % Left justify columns in tables
\usepackage{ragged2e}                                        % Left justify columns in tables
\usepackage[font=small, labelfont=bf]{caption}     % Makes all the captions small and all the Figure numbers bold
\usepackage{multirow}        				   % Allows for multiline headings and headings that span multiple columns
\usepackage{floatrow}                                           % Should allow me to position the side caption where I want it 
\usepackage{arydshln}					   % Allows for dashed lines to be inserted in tables
\usepackage{bigstrut}					   % Allows lines to be thrown before and after contents of a cell in a table, effectively vertically centering them
\usepackage{float}						   % Allows figures to be positioned at exact points in the text

\graphicspath{ {./Figures/} }               %----- Set default location of figures
\numberwithin{equation}{section}      %----- Numbers equations by section 
\usepackage{longtable}                     %----- Enables large table to be split over two pages

\newcommand{\Msun}{{\,M}$_\odot$}

\newcommand{\Dsun}{{\,D}$_\odot$}

\newcommand{\kms}{{\,\rm {km s$^{-1}$}}}

\newcommand{\alfab}{{\,[}$\alpha$/Fe{]}}

\usepackage{comment}
\usepackage[autostyle]{csquotes}

\title[The nature of the Eastern Extent]{The nature of the Eastern Extent in the outer halo of M31}

\author[J. Preston et al]   
{Janet Preston,$^1$
\ Michelle Collins,$^1$
\ R. Michael Rich$^2$
\ Rodrigo Ibata,$^3$
\newauthor \ Nicolas F. Martin$^{3,4}$
 \ Mark Fardal$^5$
\\
\\
$^1$Department of Physics, University of Surrey, Guildford, GU2 7XH, Surrey, UK. \thanks{j.preston@surrey.ac.uk} \\
$^2$Department of Physics and Astronomy, UCLA, 430 Portola Plaza, Box 951547, Los Angeles, CA 90095-1547, USA \\
$^3$Observatoire de Strasbourg, 11, rue de l'Universit\'{e}, F-67000, Strasbourg \\
$^4$Max-Planck-Institut f{\"u}r Astronomie, K{\"o}nigstuhl 17, D-69117 Heidelberg, Germany \\ 
$^{5}$ Space Telescope Science Institute, 3700 San Martin Drive, Baltimore, MD 21218, USA \\
}

\date{Accepted XXX. Received YYY; in original form ZZZ}

\pubyear{2021}

\begin{document}
\label{firstpage}
\pagerange{\pageref{firstpage}--\pageref{lastpage}}
\maketitle

\begin{abstract}
We present the first comprehensive spectroscopic study of the Andromeda galaxy's Eastern Extent. This $\sim$4$^{\circ}$ long filamentary structure, located 70-90 kpc from the centre of M31, lies perpendicular to Andromeda's minor axis and the Giant Stellar Stream and overlaps Stream C.  In this work, we explore the properties of the Eastern Extent to look for possible connections between it, the Giant Stellar Stream and Stream C. We present the kinematics and photometry for $\sim$50 red giant branch stars in 7 fields along the Eastern Extent. We measure the systemic velocities for these fields and find them to be $-$368 {\kms} $\lesssim$ $\textit v$ $\lesssim$ $-$331 {\kms}, with a slight velocity gradient of $-$0.51$\pm$0.21 {\kms} kpc$^{-1}$ towards the Giant Stellar Stream.  We derive the photometric metallicities for stars in the Eastern Extent, finding them to be metal-poor with values of $-$1.0 $\lesssim$ [Fe/H]$_{\rm phot}$  $\lesssim$ $-$0.7 with a $<$[Fe/H]$_{\rm phot}$$>$ $\sim$$-$0.9. We find consistent properties for the Eastern Extent, Stream B and one of the substructures in Stream C, Stream Cr, plausibly linking these features. Stream Cp and its associated globular cluster, EC4, have distinctly different properties indicative of a separate structure. When we compare the properties of the Eastern Extent to those of the Giant Stellar Stream, we find them to be consistent, albeit slightly more metal-poor, such that the Eastern Extent could plausibly comprise stars stripped from the progenitor of the Giant Stellar Stream.
\end{abstract}


%%%%%%%%%%%%%%%%% BODY OF PAPER %%%%%%%%%%%%%%%%%%

%%%%%%%%%%%%%%%%%%%%%%%%%%%%%%%%%%%%%%%%%%%%%%%%%%%%%%%%%%%%%%%%%%%%%%%%%
%%%%%%%%%%%%%%%%%%%%%%%%%%%%%%%%%%%%%%%%%%%%%%%%%%%%%%%%%%%%%%%%%%%%%%%%%
%--------------------------------------------------------------------------- Introduction ----------------------------------------------------------------------------------------------------
\section{Introduction} \label{introduction}
\graphicspath{ {Figures/} }    %----- Set default location of figures  

% Figure placed here to come out in the right place in the paper
\begin{figure*}
  	\centering
	\includegraphics[height=.45\paperheight, width=.8\paperwidth]{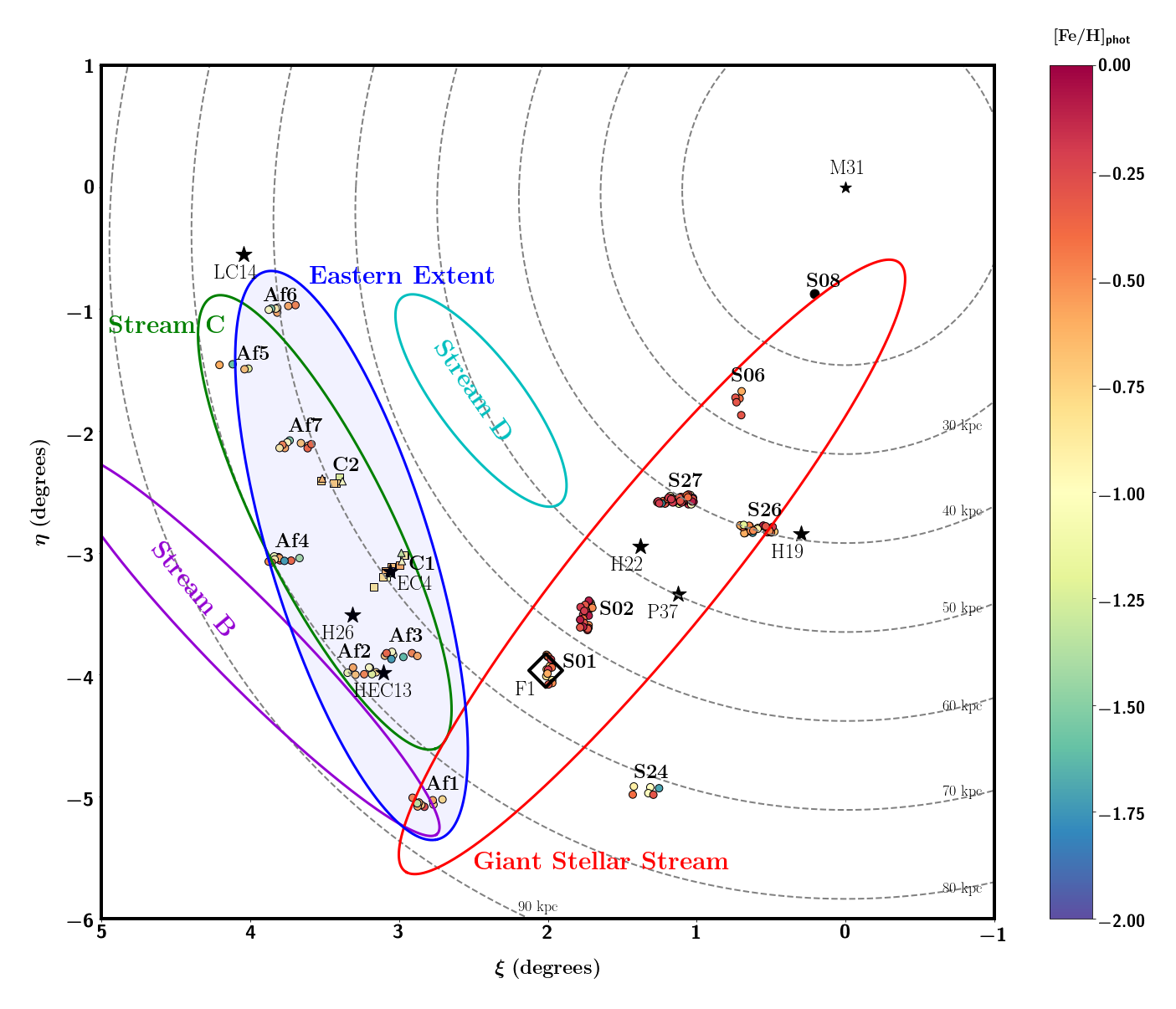}	
    	\vspace*{-4.5mm}\caption[On-sky positions of key features adjacent the EE and the GSS]
	{On-sky positions of key features adjacent the EE and the GSS colour coded by [Fe/H]$_{\rm phot}$.  The dotted lines show radial distances from M31. The icons represent the stars from fields in the EE, the GSS and Stream C, each of which is colour-coded by its [Fe/H]$_{\rm phot}$. Field S08 has no GSS stars so the plot shows only the position of the centre of the field, represented by a black circle. Stream C covers two fields, C1 and C2 both of which contain stars in the substructures Cr (denoted by square icons) and Cp (denoted by triangular icons). Stream C data taken from \cite{RefWorks:61}.  The position of the globular cluster EC4 is also shown on the plot along with the positions of other relevant globular clusters, LAMOST-C14 (labelled LC14), HEC-13, H19, H22, H26 and PAndAS-37 all of which are represented by black star shaped icons. The black open diamond, labelled F1, indicates the position of a turning point predicted by \cite{RefWorks:331}, based on N-body simulations of the GSS. The EE/GSS [Fe/H]$_{\rm phot}$ values are derived using isochrones with t = 9 Gyrs, {\alfab} = 0.0 corrected to an heliocentric distance of 845 kpc.  }
	\label{EE_Fig14}
\end{figure*}
% Figure placed here to come out in the right place in the paper
 \begin{figure*}
  	\centering
	\includegraphics[height=.35\paperheight, width=.86\paperwidth]{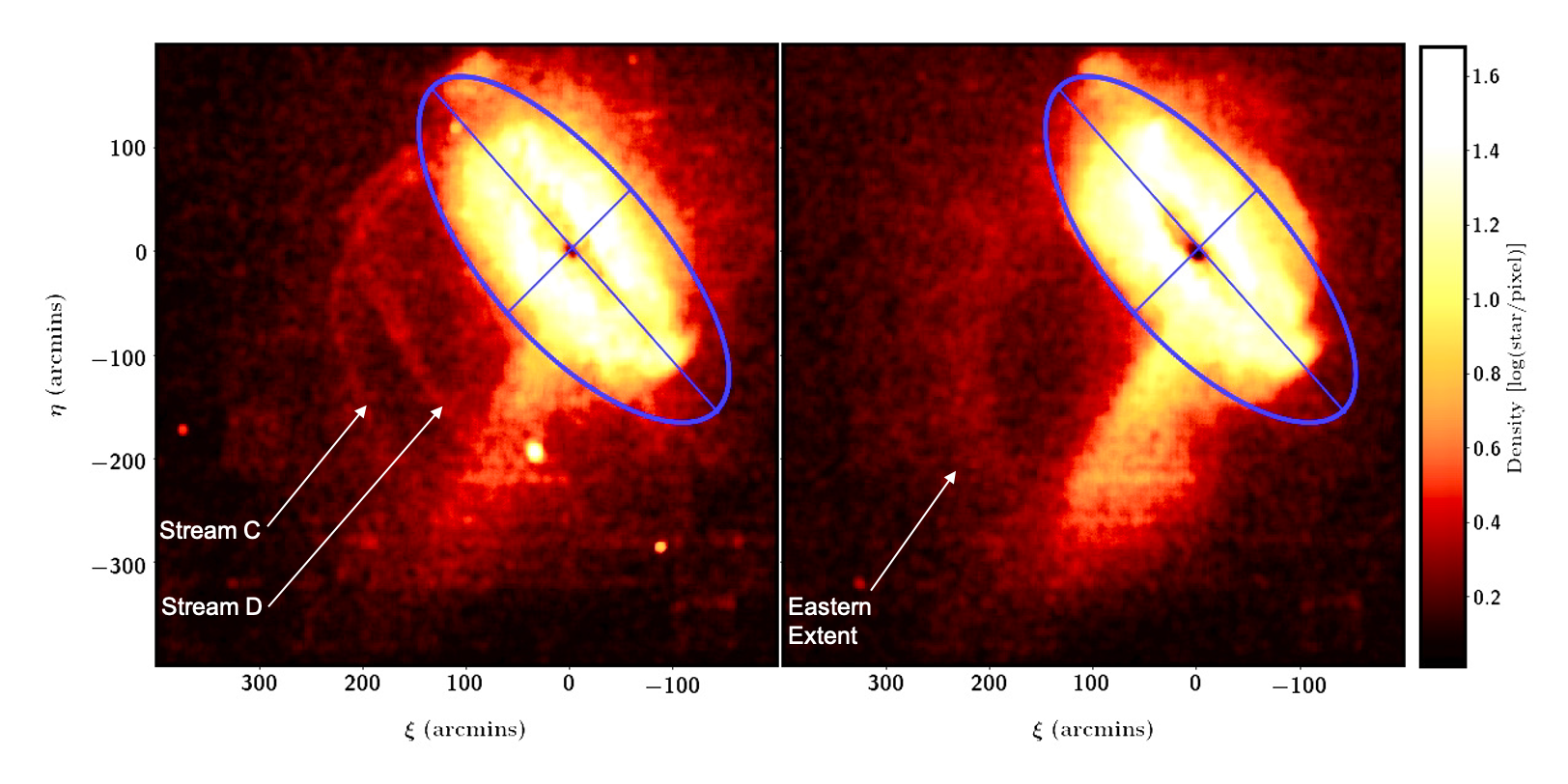}
    	\vspace*{-4.5mm}\caption[Map of M31, Giant Stellar Stream and Eastern Extent Red Giant Branch stars]
	{Map of stars in the south eastern quadrant of the M31 halo. Stars were selected from the PAndAS catalogue (\citealt{RefWorks:449}) as point source objects with 0.5 $\le$ (g-i)$_0$ $\le$ 2.5 and 21.5 $\le$ i$_0$ $\le$ 24.5. The data were convolved using a Top-hat kernel and plotted, using a log scale for the density of stars per pixel, in tangent plane coordinates centred on M31.  On both plots the M31 halo is represented by solid blue line (taking a semi-major axis of  55 kpc with a flattening of 0.6, \citealt{RefWorks:107}). The left hand panel includes stars with -2.5$\le$ [Fe/H] $\le$ -1.0 and clearly shows the narrow metal poor tracks of Streams C and D.  The right hand panel includes stars with -0.7$\le$ [Fe/H] $\le$ -0.3 and indicates the presence of a much broader metal rich feature, the Eastern Extent, that appears to trace a similar path to Stream C.  
	 }
	\label{EE_Fig1}
\end{figure*}

Evidence for intergalactic collisions and mergers can be found in the tidal debris wrapped around many galaxies in the Local Group and beyond.  Stellar streams and concentric shell systems lay testament to the destruction and accretion of smaller galaxies by larger ones (\citealt{RefWorks:513}, \citealt{RefWorks:514}, \citealt{RefWorks:515}).  These features present us with a myriad of insights into the formation and structure of their host galaxies.  If we assume that the debris within a stellar stream follows the orbit of its progenitor we can, using Newton's Law of Universal Gravitation, constrain the gravitational potential and mass of the host (\citealt{RefWorks:423, RefWorks:176}, \citealt{RefWorks:10},  \citealt{RefWorks:421}, \citealt{RefWorks:327}, \citealt{RefWorks:245}, \citealt{RefWorks:422},  \citealt{RefWorks:82}, \citealt{RefWorks:424}).

Surveys such as the Dark Energy Survey (DES,  \citealt{RefWorks:420}, \citealt{RefWorks:411}), the Sloan Digital Sky Survey (SDSS, \citealt{RefWorks:419}), the Pan-STARRS1 3$\pi$ Survey (PS1, \citealt{RefWorks:562}) and the ESA/Gaia survey (\citealt{RefWorks:204}) have discovered more than 60 streams around the Milky Way (MW) (\citealt{RefWorks:406}, \citealt{RefWorks:411},  \citealt{RefWorks:468} and \citealt{RefWorks:545}).  Further afield, the Pan-Andromeda Archaeological Survey (PAndAS, \citealt{RefWorks:58}) has led to the discovery of more than 10 streams around the Andromeda galaxy (M31) (\citealt{RefWorks:47}, \citealt{RefWorks:82}, \citealt{RefWorks:107}, \citealt{RefWorks:449}).

Of these, the Giant Stellar Stream (GSS, aka the Giant Southern Stream) in the M31 halo, is one of the most spectacular. Discovered in 2001 by \citeauthor{RefWorks:143}, using the 2.5 m Isaac Newton Telescope (INT), its kinematic and spectroscopic properties have been well explored by: \cite{RefWorks:148}, \cite{RefWorks:60, RefWorks:75},  \cite{RefWorks:444, RefWorks:82},  \cite{RefWorks:10}, \cite{RefWorks:483},  \cite{RefWorks:550}, \cite{RefWorks:447}, \cite{RefWorks:11, RefWorks:36, RefWorks:35, RefWorks:397, RefWorks:624},  \cite{RefWorks:203},  \cite{RefWorks:111}, \cite{RefWorks:495}.  Work by \cite{RefWorks:38}, discovered that the GSS had multiple stellar populations, with a metal-rich "core" ([Fe/H] $\sim$ $-$0.5) and a metal-poor "envelope" ([Fe/H] $\sim$ $-$1.3).  

The formation history of the GSS has also been explored to ascertain if it is a stellar stream or part of a more extensive shell system.  Streams are formed when stars escape from a satellite galaxy experiencing tidal disruption by a host. The stars leave the satellite through the Lagrange points of the combined host/satellite system.  Stars leaving via the inner Lagrange point (between the host and the satellite) fall into lower energy orbits with a shorter period than the satellite.  These stars form the \enquote{leading} tail of the stream.  Stars leaving via the outer Lagrange point (on the opposite side of the satellite to the host) inhabit higher energy orbits with longer periods than the satellite and form the \enquote{trailing} tail of the stream (\citealt{RefWorks:596}). 

Shells, by contrast, are open, concentric, arcs of stellar over-densities with clearly defined \enquote{edges} and exhibit characteristic light distributions such as those found around the elliptical galaxies NGC 1316 (Fornax A) and NGC 5128 (Centaurus A) (\citealt{RefWorks:598}). They comprise debris from satellite galaxies where the stripped stars have accumulated at the apocentres of their orbits (\citealt{M:780}, \citealt{RefWorks:596}, \citealt{RefWorks:598}). Shells can be exacting to detect due to their low surface brightness ($\lesssim$ 28 mag arcsec$^{-2}$), irregular morphologies and propensity to extend out to \mbox{$\gtrsim$ 100 kpc } from the galactic centre of their host. Prevalent in accretion events involving higher mass galaxies (> 10$^{12}$ {\Msun}) $\sim$4-8 Gys ago (\citealt{RefWorks:598}) shells are thought to have a number of different formation mechanisms including: the satellite approaching the host along a radial orbit (\citealt{RefWorks:715}, \citealt{RefWorks:245}, \citealt{RefWorks:598}); mergers between two low mass disk galaxies (\citealt{RefWorks:649}) and  major mergers (\citealt{M:784}, \citealt{M:779}). Studies to determine which scenario created any given shell structure, by \cite{M:778},  \cite{M:781} and \cite{M:782}, showed that shallow metallicity gradients across a shell system (with the outermost shells having the lowest metallicity) were indicative of major mergers. \cite{RefWorks:82} detected a metallicity gradient across the M31 halo that becomes increasingly metal-poor from $\langle$[Fe/H]$_{\rm phot}$$\rangle$ = $-$0.7 at R$_{\rm proj}$ = 30 kpc to $\langle$[Fe/H]$_{\rm phot}$$\rangle$ = $-$1.5 at R$_{\rm proj}$ = 150 kpc. \cite{RefWorks:35} obtained similar results with their detection of a metallicity gradient of -0.01 dex kpc$^{-1}$ between projected radii of 10 kpc and 90 kpc.  Both studies ascribed the results to M31 having undergone a massive merger during its formation history. 

N-body simulations and other models of the GSS by \cite{RefWorks:351, RefWorks:104, RefWorks:342, RefWorks:346, RefWorks:245}, \cite{RefWorks:554}, \cite{RefWorks:331}, \cite{RefWorks:447}, \cite{RefWorks:477}, \cite{RefWorks:154, RefWorks:413} , \cite{RefWorks:163}, \cite{RefWorks:109},  \cite{RefWorks:470, RefWorks:637} reproduced stream like features with properties similar to those of the GSS, while other works explored the connections between the GSS and other substructures in the M31 halo. \cite{RefWorks:147, RefWorks:446} showed a plausible association between the GSS and the North East Shelf, while \cite{RefWorks:104} reported an association between the GSS and the Western Shelf, with the possibility that they, and the North East Shelf, originated from the same progenitor.  Indeed models produced by \cite{RefWorks:155} indicated that the Western Shelf was possibly a shell created from the same debris that produced the GSS, most likely during the third orbital wrap of a $\sim$10$^9${\Msun} progenitor around M31. 

Other features such as Streams B, C and D, which lie perpendicular to the GSS, were first reported by \cite{RefWorks:38}, who discounted them being associated with the GSS given their very different stellar populations. This could account for why few models of the GSS reproduce their structures, although \cite{RefWorks:342} did develop a model that produced \enquote{curious arcs} that qualitatively resembled Streams C and D.  In this model the kinematics matched the observational data but at distances much further from the centre of M31 than those for the actual features and the metallicities were more metal-poor than the GSS.  With the progenitor modelled as a strongly rotating disk galaxy, \citeauthor{RefWorks:342} postulated that the motion of the disk caused lateral movement of some of the debris thus giving rise to the features resembling Streams C and D.

% Table placed here to appear in the right place in the text
\begin{table*}
	\centering
	\setlength\extrarowheight{2pt}
	\caption[Properties for observed fields in the Eastern Extent and the Giant Stellar Stream]
	{Properties for observed fields in the EE and GSS including: field name; date observations were made; observing PI; Right Ascension and Declination of the centre of each field; projected distance of the centre of the field from M31(D$_{M31}$) and the number of stars likely to belong to each of the stellar populations (i.e. EE/GSS, M31 and the MW) based on their radial velocities. The $\alpha$ and $\delta$ for the centre of each field are determined by taking the mean of the $\alpha$s and $\delta$s for all stars on that field.}		
	\label{EE_table:1}
	\begin{tabular}{ccccccccc} 
		\hline
		\multirow{2}{*} {Field} & 
		\multirow{2}{*} {Date} &
		\multirow{2}{*} {PI} &
		\multirow{2}{*} {$\alpha_{\rm J2000}$} & 
		\multirow{2}{*} {$\delta_{\rm J2000}$}  & 
		\multicolumn{1}{c}{$D_{M31}$}  &     
		\multicolumn{3}{c}{No. of candidate stars within...}  \\ [0.5ex] 
		\cline{7-9}
		&  &  &  &  & kpc  & EE/GSS & M31 & MW  \\ [0.5ex]
		\hline 		
		Af1 & 2015-09-17 & Rich &   00:56:40.00  & +36:10 54.00   &   79.0  &     9  &   3  &   16\\ [0.5ex]
		Af2 & 2015-09-17 & Rich &   00:59:07.68  & +37:14:23.07   &   70.0  &     8  &   6  &   20 \\ [0.5ex]
		Af3 & 2015-09-17 & Rich &   00:57:54.45  & +37:22:33.05   &   67.0  &     8  &   7  &   14\\ [0.5ex]
		Af4 & 2016-09-04 & Rich &   01:01:57.38  & +38:07:03.49   &   67.0  &   10  &   10  &   14 \\ [0.5ex]
		Af5 & 2016-09-04 & Rich &   01:04:01.91  & +39:40:30.22   &   60.0  &    4   &   12  &   22 \\ [0.5ex]
		Af6 & 2016-09-04 & Rich &   01:02:41.88  & +40:10:06.07   &   54.0  &    7   &   11  &   23 \\ [0.5ex]
		Af7 & 2016-09-04 & Rich &   01:01:44.69  & +39:03:33.85   &   58.0  &    9  &   14  &   18\\ [0.5ex]
		
		S01 & 2002/2003/2004  & Ibata &   00:52:44.45  & +37:17:52.77   &   61.0  &    13  &   12  &   21 \\ [0.5ex]
		S02 & 2002/2003/2004  & Ibata &   00:51:33.39  & +37:44:12.71   &   54.0  &    27  &     8  &   23 \\ [0.5ex]
		S06 & 2002/2003/2004  & Ibata &   00:46:26.85  & +39:30:58.00   &   26.0  &      5  &   15  &   8\\ [0.5ex]
		S08 & 2002/2003/2004  & Ibata &   00:43:49.91  & +40:23:31.68   &   12.0  &      0  &   56  &   9\\ [0.5ex]
		S24 & 2002/2003/2004  & Ibata &   00:49:30.95  & +36:18:48.42   &   70.0  &      6  &   18  &   76 \\ [0.5ex]
		S26 & 2002/2003/2004  & Ibata &   00:45:48.17  & +38:27:43.06   &   39.0  &    23  &   30  &   65 \\ [0.5ex]
		S27 & 2002/2003/2004  & Ibata &   00:48:33.59  & +38:41:44.69   &   38.0  &    47  &   13  &   61 \\ [0.5ex]
		\hline
	\end{tabular}
\end{table*}

However, over the course of the PAndAS programme, increasingly detailed maps of M31's halo have been revealed. Initially, Streams C and D were identified as narrow, metal-poor stream features (\citealt{RefWorks:38}). But over time a much broader and more metal-rich feature overlapping Stream C became apparent in maps of this region (see Figures \ref{EE_Fig14} and \ref{EE_Fig1}, and cf. Figure 9 in \citealt{RefWorks:82}). Correspondingly, spectroscopic observations in this area revealed two distinct kinematical components with differing metallicities (Streams Cr and Cp, \citealt{RefWorks:61} and \citealt{RefWorks:11}). The broad metal-rich feature spans 70-90 kpc from the centre of M31 along M31's minor axis, and follows an arc of length $\sim$80 kpc as it approaches M31's disk. It seems to connect seamlessly with the GSS and have a similar stellar population to that stream. It is currently unclear whether this feature is stream-like in nature or even its own distinct structure. In this paper we provide the first designation of this metal-rich feature as the Eastern Extent (EE), rather than labelling it as one of M31's distinct tidal streams. Understanding more about the nature of this intriguing feature, whether or not it is associated with Stream C, and how it relates to the GSS could enhance our understanding of how they and the other striking debris structures in M31's halo were formed.

In this work we analyse the kinematic and photometric properties of stars in seven fields along the length of the EE.  We compare these with corresponding properties of the surrounding features to see if there are any possible associations between them.  We present the results of our analysis as follows: Section \ref{Observations} describes the observations and data reduction process; Section \ref{Analysing membership of the Eastern Extent and the Giant Stellar Stream} describes our approach to analysis of the data,  Section \ref{Discussion} contains a discussion of our findings and we present our conclusions in Section \ref{Conclusions}.

%%%%%%%%%%%%%%%%%%%%%%%%%%%%%%%%%%%%%%%%%%%%%%%%%%%%%%%%%%%%%%%%%%%%%%%%
%%%%%%%%%%%%%%%%%%%%%%%%%%%%%%%%%%%%%%%%%%%%%%%%%%%%%%%%%%%%%%%%%%%%%%%%
%-----------------------------------------------------------Section 2 - Observations -------------------------------------------------------------------------------------------------
\section{Observations} \label{Observations}
\graphicspath{ {Figures/} }

The dataset comprises observations in fields along the EE and GSS as shown in Figure \ref{EE_Fig14} and detailed in Table \ref{EE_table:1}. The GSS data (fields S01-S27) were obtained as a spectroscopic follow-up of the substructures around M31 previously detected by \cite{RefWorks:143}. The data were obtained over 8 nights in September 2002, 2003 and 2004, using the Keck II Telescope fitted with the DEep-Imaging Multi-Object Spectrograph (DEIMOS). Described by \cite{RefWorks:444}, the observations used the DEEP2 slit mask approach and covered wavelengths in the range  6400\AA{} - 9000\AA, with a spectral resolution $\sim$0.6\AA{}.  The data were then reduced using an early version of the DEEP2 pipeline as described by \cite{RefWorks:748}.

Data for the EE were obtained over two observing runs.  Data for the first run, including fields Af1 - Af3, were obtained during a single nights viewing, 1 hour per field (3 x 20 minute integrations), in September 2015.  From this data we were able to identify $\sim$10 member stars per field.  For the second set of observations, i.e. fields Af4-Af7, we increased the observing time to 2 hours per field (4 x 20 minute integrations) with the aim of confirming additional member stars in these more diffuse fields.  Both observing runs used the Keck II Telescope fitted with the DEIMOS and focussed on the Calcium Triplet (CaT) region located between wavelengths 8400\AA{} and 8700\AA.  Both observing runs used the OG550 filter with the 1200 lines/mm grating with a resolution $\sim$1.1\AA{} - 1.6\AA{} at full width half maximum (FWHM).  The data were reduced using the pipeline described by \cite{RefWorks:216}.  It corrected for scattered light, flat-field pixel variations and illumination in the telescope.  It also calibrated the pixel wavelengths and determined the velocity and related uncertainties by creating a model spectrum that comprised a continuum and absorption profiles of the CaT lines.  It then cross-correlated the model with non-resampled stellar data to obtain the Doppler shift and the CaT line widths before correcting the derived velocity data to the heliocentric frame.
 
For all fields our highest priority targets (which were expected to have the highest probability of membership of the EE/GSS) were bright stars lying on the EE/GSS Red Giant Branches (RGBs) i.e.  21.0 < \textit{i$_0$} < 22.5. The next priority was fainter stars on the RGB with 22.5 < \textit{i$_0$} < 23.5.  The remainder of the field was then filled with stars with 20.5 < \textit{i$_0$} < 23.5 and 0.0 < \textit{g-i} < 4.0.

Throughout this work we take the heliocentric distance of M31 to be 783 $\pm$ 25 kpc (\citealt{RefWorks:56}).

%%%%%%%%%%%%%%%%%%%%%%%%%%%%%%%%%%%%%%%%%%%%%%%%%%%%%%%%%%%%%%%%%%%%%%%
%%%%%%%%%%%%%%%%%%%%%%%%%%%%%%%%%%%%%%%%%%%%%%%%%%%%%%%%%%%%%%%%%%%%%%%
%---------------------------Section 3:  Determining membership of the Eastern and Giant Stellar Streams -------------------------------------------------------
\section{Analysing the Eastern Extent and the Giant Stellar Stream} \label{Analysing membership of the Eastern Extent and the Giant Stellar Stream}
\graphicspath{ {Figures/} } 

%------------------------------------------------------------------------------------------------------------------------------------------------------------
\subsection{Systemic Velocities} \label{Systemic Velocities}

%Figure placed here to appear in the right place in the text
\begin{figure*}
	\begin{center}	
		\vspace{-3.0mm}                        % This adjusts the amount of vertical space between the top row of the sub-plots and the top of the page
		\includegraphics[height=.52\paperheight, width=.84\paperwidth]{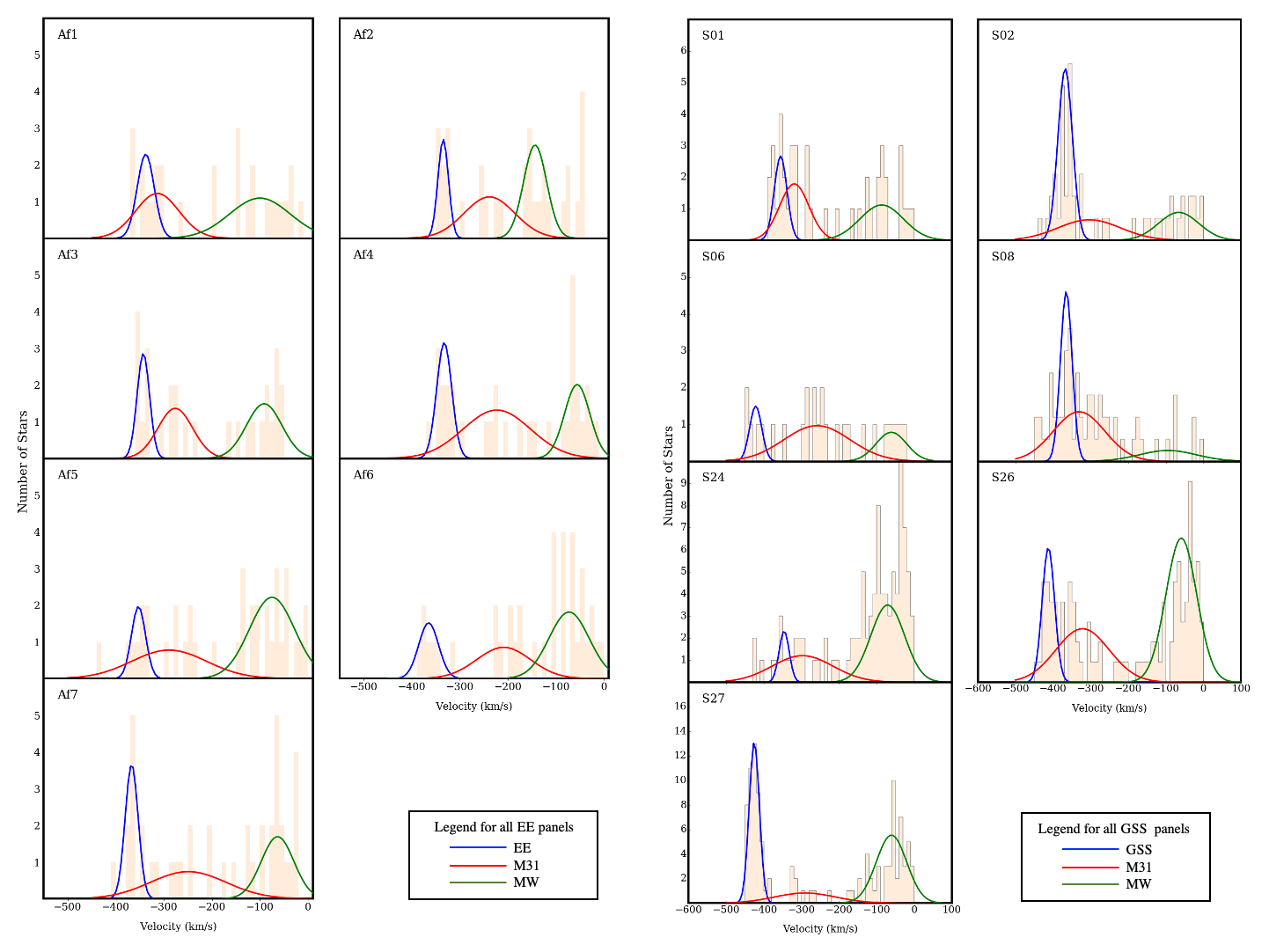}
  		\vspace*{-7mm}\caption[EE and GSS velocity histograms]
	{Kinematic analysis of EE and GSS fields showing the velocity histograms fields overlaid with membership probability distribution function for each of the three stellar populations - shown in blue for the EE or GSS, red for M31 and green for the MW.  }
	\label{EE_Fig6}
	\end{center}
\end{figure*}

Initially candidate EE/GSS stars were selected based on radial velocity, which we expected to be $\sim$$-$355 {\kms} (\citealt{RefWorks:11}). Figure \ref{EE_Fig6} shows the velocity distribution for the stars in each field overlaid with probability distribution function (pdf) derived from a Monte Carlo Markov Chain (MCMC) analysis.  
With bin sizes selected to provide an optimal representation of the data structures and to resolve even the coldest stellar populations present in our data, these figures show clear peaks around the EE/GSS velocity as well as two peaks for other candidate stellar populations i.e. MW stars (v$_r$ $\sim$ $-$80 {\kms}, \citealt{RefWorks:42}) and M31 halo stars (v$_r$ $\sim$ $-$300 {\kms}, \citealt{RefWorks:444}). To assign a star to a particular stellar population we define a Gaussian pdf for each:
\begin{equation} 
	\label{EE_eq:3MW}
	\begin{multlined}
		P_{\rm pop}  = \frac{1}{\sqrt{ 2 \pi( \sigma_{v,\rm pop}^2 + v^2_{\rm err,i})}} \times \mathrm{exp}\Bigg[-\frac{1}{2} 
		\bigg( \frac{v_{r,\rm pop} - v_{r,i}} {\sqrt{\sigma_{v,\rm pop}^2 + v^2_{\rm err,i}}}
		\bigg)^2 
		\Bigg]
	\end{multlined}
\end{equation}
where: $v_{r,i}$ is the velocity of a given star with an uncertainty of $v_{\rm err,i}$ and $P$$_{\rm pop}$, $v$$_{\rm pop}$ and $\sigma$$_{v,\rm pop}$ are the resulting pdf, systemic velocity and velocity dispersion, respectively, for the EE/GSS, M31 and MW stellar populations. 

The likelihood function for membership of the EE or GSS, based on velocity, is defined as:
\begin{equation} 
	\label{EE_eq:5}
	\begin{multlined}
		\mathrm{log}[\mathcal{L}(v_{\rm pop}, \sigma_{v,\rm pop})] = \sum_{i=1}^{N} \mathrm{log}(\eta_{\rm M31} P_{i,\rm M31} +\\
		 \shoveleft[2.5cm] \eta_{\rm MW} P_{i,\rm MW} + \eta_{\rm feat} P_{i,\rm feat})
	\end{multlined}
\end{equation}
\\
where $P$$_{\rm feat}$,  $P$$_{\rm M31}$ and $P$$_{\rm MW}$ are the resulting pdfs for the EE/GSS, M31 and MW stellar populations and  $\eta_{\rm M31}$, $\eta_{\rm MW}$ and $\eta_{\rm feat}$ are the fraction of stars within each stellar population (where $\eta_{\rm feat}$ represents either the EE or GSS depending on context) and:
\begin{equation} 
	\label{EE_eq:5a}
	\eta_{\rm feat} = 1 - (\eta_{\rm M31}  + \eta_{\rm MW})
\end{equation}

We then use {\sc emcee} (\citealt{RefWorks:571}, \citealt{RefWorks:63}) to fit the Gaussians simultaneously and derive the posterior distribution for the systemic velocity, velocity dispersion and fraction parameters for the stellar populations in each field. Our set up for this algorithm includes:
\begin{itemize}
	\item Selecting stars where $-$450.0 $\leq$  $v_{r,i}$  {\kms} $\leq$ 0.0 and the velocity uncertainty is $\leq$ 20 {\kms}.
	 \item Setting the initial velocity value for the EE/GSS to \mbox{$\textit v_r$ = $-$350 {\kms}}.  
	 \item Obtaining initial values for the velocity dispersions with respect to the distance of the centre of the field from M31 (see \citealt {RefWorks:10} and \citealt {RefWorks:235}) using Equation \ref{EE_eq:4}.
	\item Basing initial values for the fraction parameters: $\eta_{\rm M31}$; $\eta_{\rm MW}$ and $\eta_{\rm feat}$ on the velocity distribution for each field as seen in Figure \ref{EE_Fig6}. 
	\item Defining the priors as shown in Table \ref{EE_table:7a}.
\end{itemize}

\begin{equation} 
	\label{EE_eq:4}
	\sigma_{v}(R) = \bigg(152 - 0.9 \frac{R}{1\: \rm{kpc}}\bigg)  \: \rm km s^{-1}  kpc^{-1}
\end{equation}

\begin{table}
	\centering
	\setlength\extrarowheight{2pt}	
	\caption[Priors for the {\sc emcee} analysis for the EE and GSS ]
		{Priors for the {\sc emcee} analysis. (a) dimensions apply to all parameters and priors except for the fraction parameters which are dimensionless. (b) for fields Af6, S02, S06, S08, S26 and S27, the prior for M31 is defined as 0 $\le$ $\sigma$$_v$ $\le$ 150.0 {\kms} as the velocity dispersions, at the distance of these fields from M31, are found to be > 100 {\kms}. }
		\label{EE_table:7a}
	\begin{tabular}{ll} 
		\hline
	 	Parameter ({\kms})$^a$   & Prior ({\kms})$^a$   \\
		\hline
		v$_{r, \rm feat}$                     &   $-$450 $\le$ $v_{r, \rm feat}$ $\le$ $-$300 \\
		v$_{r, \rm M31}$                    &   $-$400 $\le$ $v_{r, \rm M31}$ $\le$ $-$200 \\		
		v$_{r, \rm MW}$                     &   $-$150 $\le$ $v_{r, \rm MW}$  $\le$ 50 \\		
		$\sigma_{v, \rm feat}$            &    0 $\le$ $\sigma$$_{v, \rm feat}$ $\le$ 20\\
		$\sigma_{v, \rm M31}$$^b$   &    0 $\le$ $\sigma$$_{v, \rm M31}$ $\le$ 100\\
		$\sigma_{v, \rm MW}$           &    0 $\le$ $\sigma$$_{v, \rm MW}$ $\le$ 150\\
		\multirow{ 2}{*}{$\eta$}          &    0 $\le$ $\eta$ $\le$ 1 with \\
		                                              &    $\eta_{\rm feat}$ + $\eta_{\rm M31}$ + $\eta_{\rm MW}$  = 1\\[0.3ex]
		\hline
	\end{tabular}
\end{table}

 % Table placed here to appear in the right place in the text
 \begin{table}
	\centering
	\setlength\extrarowheight{2pt}	
	\caption[Results of the kinematic analysis of Eastern Extent and Giant Stellar Stream fields]
		{Results of the kinematic analysis of EE and GSS fields. The table includes the number of confirmed stars in the EE or GSS stellar population in each field.}
		\label{EE_table:7}
	\begin{tabular}{lccc} 
		\hline
	 	Field & v$_r$ & $\sigma_v$ & Confirmed\\
		 & {\kms}& {\kms}  & stars\\
		\hline
Af1 &   $-$337.7 $^{+ 11.4 }_{- 12.9 }$   &   15.5 $^{+ 3.2 }_{- 6.5 }$      &  9   \\[1.4ex]
Af2 &   $-$334.8 $^{+ 3.3 }_{- 3.2 }$       &   4.9 $^{+ 5.0 }_{- 3.3 }$       &   8   \\[1.4ex]
Af3 &   $-$340.7 $^{+ 5.3 }_{- 5.5 }$       &   9.6 $^{+ 5.8 }_{- 5.1 }$       &   8   \\[1.4ex]
Af4 &   $-$332.5 $^{+ 6.0 }_{- 6.4 }$       &   14.0 $^{+ 3.9 }_{- 5.7 }$      & 10  \\[1.4ex]
Af5 &   $-$352.9 $^{+ 15.8 }_{- 15.2 }$   &   12.4 $^{+ 5.2 }_{- 7.0 }$      &  4   \\[1.4ex]
Af6 &   $-$365.0 $^{+ 10.7 }_{-8.8 }$     &   19.4 $^{+ 12.2 }_{-8.8 }$      &  6  \\[1.4ex]
Af7 &   $-$367.0 $^{+ 6.1 }_{- 4.8 }$       &   10.1 $^{+ 6.7 }_{- 7.2 }$      &   9  \\[1.4ex]

S01 &   $-$353.1 $^{+ 9.8 }_{- 8.5 }$      &   14.9 $^{+ 3.7 }_{- 8.4 }$     &    13    \\[1.4ex]
S02 &   $-$369.0 $^{+ 4.1 }_{- 4.3 }$      &   17.4 $^{+ 1.8 }_{- 3.2 }$     &    27   \\[1.4ex]
S06 &   $-$431.1 $^{+ 12.6 }_{- 11.4 }$   &   13.8 $^{+ 4.4 }_{- 9.0 }$     &     5   \\[1.4ex]
S24 &   $-$346.6 $^{+ 8.2 }_{- 6.1 }$      &   8.9 $^{+ 7.6 }_{- 6.0 }$       &      6  \\[1.4ex]
S26 &   $-$410.7 $^{+ 4.9 }_{- 5.8 }$      &   16.1 $^{+ 2.7 }_{- 4.1 }$     &    23  \\[1.4ex]
S27 &   $-$426.1 $^{+ 1.7 }_{- 1.7 }$      &   10.8 $^{+ 1.6 }_{- 1.4 }$     &    47  \\[1.4ex]
\hline
	\end{tabular}
\end{table}

Along with our results (see Table \ref{EE_table:7}) we see that with an acceptance fraction $\sim$0.3 (which is in the range 0.2 - 0.5 recommended by \citealt {RefWorks:382}) we have a statistically valid number of independent samples to represent the data.   We are also satisfied that, with a precision (i.e. the square root of the number of independent samples) $\sim$0.003 that is very much smaller than the posterior uncertainties, the MCMC chains have converged.

We note that for some of the fields, i.e. Af1, Af2, S01 and S08, the MW is not well represented by a single Gaussian.  We consider fitting multiple Gaussians to obtain a better model for these data. However, for fields Af1 and S01 we decide that this would overfit the data and not enhance the quality of the results.  Looking at the data for field S08, we see that there are so few stars in the MW area of the histogram that fitting more than one Gaussian would entail trying to obtain meaningful constraints from one or two stars at best. Given this is statistically unsound we, again, decide not to proceed any further and accept the results obtained from the original analysis. For field Af2, we do fit two Gaussians to the candidate MW stars to see what impact this has on the results for M31 and the EE.  We find that the data can be well represented with two Gaussians centred around $-$50{\kms} and $-$150{\kms}. However, this has negligible effect on the posterior values obtained for the M31 and EE stellar populations.  

To determine the most appropriate model to adopt for further analysis we compare them using the extended Akaike information criterion, AIC$_c$ (for use with small datasets) and the Bayesian information criterion, BIC.  We use equations defined by \cite{RefWorks:527}:
\begin{align}
	\label{EE_eq:18}
	AIC_c = -2{\rm log}(\mathcal{L})+ 2K + \frac{2K(K+1)}{n-K-1} 
\end{align}
and:
\begin{align}
	\label{EE_eq:19}
	BIC = -2{\rm ln}(\mathcal{L}) + K
\end{align}
where: $\mathcal{L}$ is the maximum value of the likelihood function for a given model, K is the number of parameters to be estimated and n is the number of data points in the analysis (in our case, the number of stars in the field).  Neither the AIC$_c$ nor the BIC results provide any insights into the absolute quality of either model, they merely indicate the quality of one relative to the other.  The model with the lowest AIC$_c$ or BIC is considered to be the optimum representation of the data. In our case the model with the single Gaussian fit has the lower scores for both the AIC$_c$ and BIC so we adopt the results from this model for further analyses and inclusion in the paper. 

Having obtained a Gaussian posterior distribution function for each of the three stellar populations, we derive the probabilities for each star belonging to a given population using:
\begin{equation} 
	\label{EE_eq:6}
	P_{\rm vel} = \frac{P_{\rm feat}}{P_{\rm M31} + P_{\rm MW} + P_{\rm feat}}
\end{equation}
\\
with the probability of being a contaminant given by: 
\begin{equation} 
	\label{EE_eq:8}
	P_{\rm contam} = \frac{P_{\rm M31} +  P_{\rm MW}} {P_{\rm M31} +  P_{\rm MW} + P_{\rm feat}}
\end{equation}

To further refine the stellar populations we overlay the RGBs of the EE and GSS with an array of isochrones with $-$2.0 $\le$ [Fe/H] $\le$ 0.0, following the approach by  \cite{RefWorks:38} and \cite{RefWorks:11}. Using the Dartmouth Stellar Evolution Database (\citealt{RefWorks:141}), we generate isochrones, prepared for the CFHT-MegaCam ugriz filter, aged 9 Gyrs (\citealt{RefWorks:552}) and {\alfab} = 0.0 to form our array.  We correct the isochrones for reddening and distance, for which we use a value 845 kpc. This heliocentric distance is based on data from \cite{RefWorks:148}, who ascertained that the GSS, in places, lies up to 100 kpc behind M31. They determined distances to 8 fields to the south-east of M31 that are very closely aligned with, and cover the full range of, the GSS fields, so we take the average of the distances to these fields to correct the isochrones.  This distance is also consistent with the average distance of the 24 GSS fields analysed by \cite{RefWorks:111} within 90\% confidence limits. We also use this distance to correct the isochrones for our EE analysis, believing it to be appropriate in light of our hypothesis that the EE comprises stars stripped from the GSS's progenitor.

\begin{figure*}
	\begin{center}
		\includegraphics[height=.3\paperheight, width=.8\paperwidth]{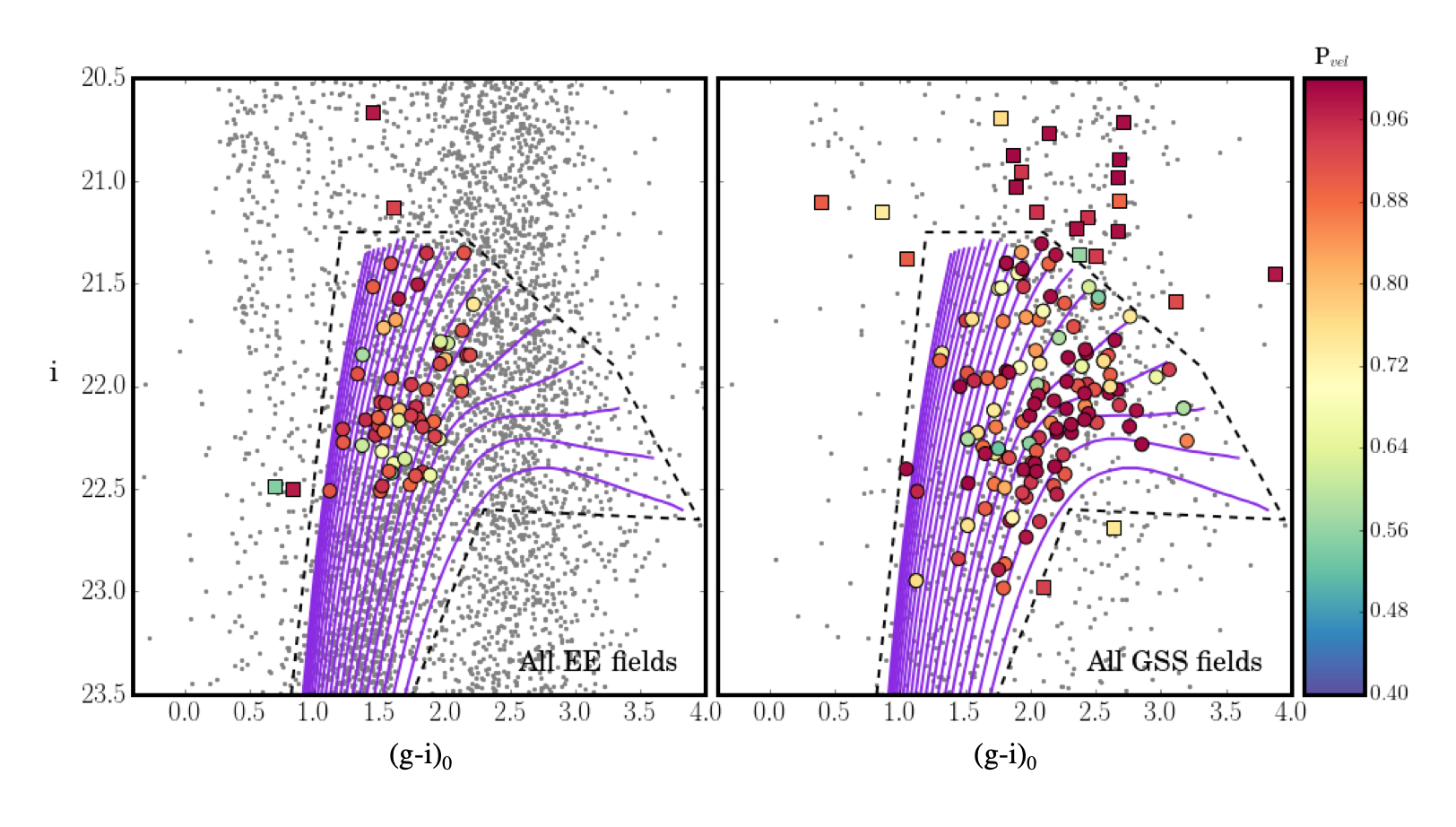}
		\vspace*{-9mm}\caption[CMD for Eastern Extent and Giant Stellar Stream fields]
	{CMD for EE and GSS fields with an extinction and distance ({\Dsun} = 845 kpc) corrected array of isochrones aged 9 Gyrs,  {\alfab} = 0.0 and metallicities of $-$2.0 $\le$ [Fe/H] $\le$ 0.0. The small black dots show stars from the main PAndAS catalogue that lie within 20 arcmins of one of the fields in each feature (AF7 for the EE and S27 for the GSS). The stars are colour coded by their strength of association with their nearest isochrone. The dashed line indicates the limits of the bounding box.  Stars outside the box are excluded from the stellar populations and further analysis.}
	\label{EE_Fig3}
	\end{center}
\end{figure*}

We surround the isochrone grid with a bounding box and plot the stars that have a high probability (P$_{\rm vel}$ $\ge$ 50\%) of being members of the EE/GSS.  Stars within the bounding box are likely to be not only EE/GSS candidates but also M31 halo stars.  So while we cannot state definitively that stars within the box are members of the EE/GSS we are confident that stars outside the box, lying further away from the EE/GSS RGBs, are unlikely to be members of these structures (see Figure \ref{EE_Fig3}) so we exclude them from all further analysis.

\begin{figure}
  	\centering
	\includegraphics[width=\columnwidth]{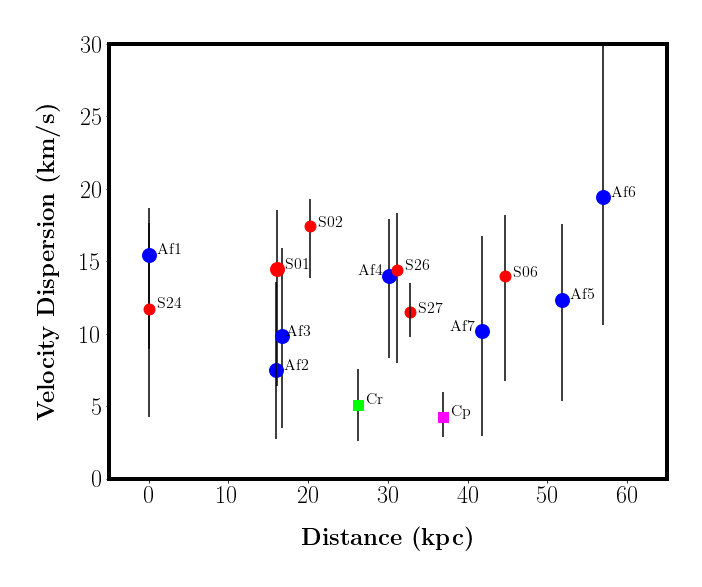}
    	\vspace*{-9mm}\caption[Velocity Dispersion for EE and GSS fields]
	{Velocity dispersions for the EE (blue icons) and the GSS (red icons). The velocity dispersions and error bars are obtained from the {\sc emcee} algorithm (\citealt{RefWorks:63}).  The plot also shows the velocity dispersion for Stream Cr (square green icon) and Stream Cp (square magenta icon), both obtained by taking the average velocity dispersion for the respective substructures as recorded by \cite{RefWorks:61}. The field locations are obtained using the mean value of all the $\alpha$s and $\delta$s for all stars in each respective field. The distances are measured from field Af1 for the EE, Stream Cr and Stream Cp and from field S24 for the GSS. }
	\label{EE_Fig77}
\end{figure}

We then analyse the velocity dispersions of the EE and GSS confirmed stellar populations.  First we plot the velocity dispersion for each field, as calculated by the {\sc emcee} algorithm, and see that they all lie within the range \mbox{4 {\kms}  $\lesssim$ $\sigma_v$ $\lesssim$ 20 {\kms}}, see Figure \ref{EE_Fig77}.

\begin{figure}
  	\centering
	\includegraphics[width=\columnwidth]{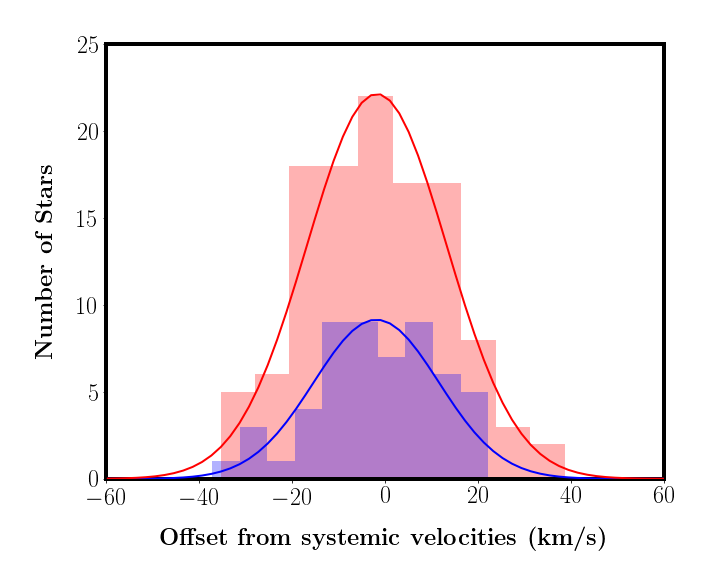}
    	\vspace*{-9mm}\caption[Distribution of velocity dispersions for EE and GSS fields]
	{Distribution of velocities with respect to the systemic velocity of fields in the EE  (blue) and GSS (red).  Both features have similar narrow peaks indicating velocity dispersions of \mbox{$\sim$14.7 {\kms}} for the EE and \mbox{$\sim$ 15.0 {\kms} }for the GSS. }
	\label{EE_Fig78}
\end{figure}

We also plot the velocity distributions.  We shift the values in each feature to a common frame by subtracting the mean value for the field from the velocity of each star in the confirmed EE/GSS stellar populations. Figure \ref{EE_Fig78} shows that both the EE and the GSS have similar velocity distributions with narrow peaks centred around $\sigma_{v \mathrm EE}$ = \mbox{13.5 {\kms}} and $\sigma_{v \mathrm GSS}$ = 15.0 {\kms}. These findings are consistent with those of \cite{RefWorks:176},  \cite{RefWorks:483},  \cite{RefWorks:351}, \cite{RefWorks:331}, \cite{RefWorks:550} and  \cite{RefWorks:11, RefWorks:624}.  They are also consitent with the progenitor of the stream being a low mass dwarf galaxy 10$^9$ $\lesssim$ {\Msun} $\lesssim$ 10$^{10}$ (\citealt{RefWorks:176}, \citealt{RefWorks:477} and \citealt{RefWorks:245}).

%------------------------------------------------------------------------------------------------------------------------------------------------------------
\subsection{Photometry}\label{Metallicities}
We examine the spectra for the EE fields and find most have S/N $<$ 3.  This means that derived spectroscopic metallicities are likely to be unreliable.  To obtain a reasonable estimate of the $\langle$[Fe/H]$_{\rm spec}$$\rangle$ per field, we stack the spectra following the approach adopted by \cite{RefWorks:444}, \cite{RefWorks:160, RefWorks:175},  \cite{RefWorks:447} and \cite{RefWorks:52, RefWorks:171}.  

Using the approach outlined by \cite{RefWorks:42}, we prepare the individual spectra for the EE candidate stars by correcting for their stellar velocities, smoothing and normalising them using a median filter, weighting them by their S/N, interpolating to a common wavelength then co-adding their fluxes.  We fit a continuum and the CaT lines of the co-added spectrum  simultaneously to obtain their equivalent widths.  As documented by \cite{RefWorks:318}, \cite{RefWorks:610},  \cite{RefWorks:272}, there is a well established, calibrated, relationship between the equivalent widths of the CaT lines and the [Fe/H]$_{\rm spec}$.  Ideally, we would use all three of the CaT lines, however not all of the co-added spectra have all three clearly defined.  In some instances the first line is contaminated by sky-lines so we adopt the metallicity estimator from \cite{RefWorks:272} i.e.:

\begin{equation} 
	\label{eq:21}
	\begin{multlined}
		[\mathrm{Fe/H]} = a + bM + cEW_{(2+3)} + dEW_{(2+3)}^{-1.5} + eEW_{(2+3)}M   
	\end{multlined}
\end{equation}
\\
where: $a$, $b$, $c$, $d$ and $e$ are taken from the calibration to the Johnson-Cousins \textit{M$_I$} values and equal to -2.78, 0.193, 0.442, -0.834 and 0.0017 respectively; and EW$_2$ and EW$_3$ are the the equivalent widths for the CaT lines at 8542\AA{} and  8662\AA {} respectively. EW$_{(2+3)}$ = EW$_2$ and EW$_3$. $M$ is the absolute magnitude of the star given by: 
\begin{equation} 
	\label{eq:22}
	M = i_j - 5 \times \mathrm{log}{_{10}}(\mathrm{D}{_{\odot}}) + 5
\end{equation}	
where: $i_j$ is the i-magnitude of the star and $D{_{\odot}}$ is the heliocentric distance for the star, which we assume to be 845 kpc for all stars.  We note that the PAndAS i-band is based on the Vega system and will, therefore, need to be transformed to the Johnson-Cousins system to determine [Fe/H]$_{\rm spec}$ defined in equation \ref{eq:22}.  The transformation is given by:
\begin{equation} 
	\label{eq:32}
	i_j = i - 0.08 \times (g-i) +0.06
\end{equation}	
where: $\textit{i}$ and $\textit{g}$ are the i-band and g-band values for the star. Uncertainties on the metallicity are determined by combining in quadrature the uncertainties on the equivalent widths, obtained from the covariance matrix produced by the fitting process.  We present our results, which show the EE has $-$1.3 $\lesssim$  [Fe/H]$_{\rm spec}$ $\lesssim$ $-$0.5, in Table \ref{EE_table:20}.

We then undertake an analysis of the Na II doublet lines in the spectra to see if we can further refine the stellar population by identifying and removing any MW dwarf star contaminants as done by  \cite{RefWorks:192}. However, the lines are barely discernible in the low S/N spectra and yield unreliable results, so they are not included here.

Our next step would be to perform the same analysis on the spectra of the GSS fields.  However, the extracted spectra are no longer available. Due to their peculiar mask design these older observations cannot be reduced with the \cite{RefWorks:216} software without significant recoding so we are unable to compare the EE and GSS spectroscopic metallicities. However, in order to undertake some form of metallicity comparison, we determined the photometric metallicities for the stars in both stellar populations based on isochrone proximity. 

Using the isochrone grid described earlier,  we match the stars in each field to the nearest isochrone and set their [Fe/H]$_{\rm phot}$ to that of the isochrone. To determine the uncertainties on these values we take into account that EE and GSS stellar populations have a variety of distances, ages and $\alpha$-element abundances and repeat our [Fe/H]$_{\rm phot}$ analysis using isochrones with the same metallicity ranges for:
\begin{itemize}
	\item \textit{change in distance}: age =  9 Gyrs, heliocentric distance = 783 kpc.  When we compare this with our original analysis, we find our results are shifted by +0.1 dex for both the EE and the GSS.
	\item \textit{change in alpha-enrichment}: age =  9 Gyrs, heliocentric distance = 845 kpc and {\alfab} = 0.2.  We find this shifts our results by $-$0.12 dex for the EE and by $-$0.13 dex for the GSS.
	\item \textit{change in age}: age = 12 Gyrs, heliocentric distance = 845 kpc and {\alfab} = 0.0.  This shifts our results by $-$0.1 dex for both features.
\end{itemize}

\begin{figure}
  	\centering
	\includegraphics[width=\columnwidth]{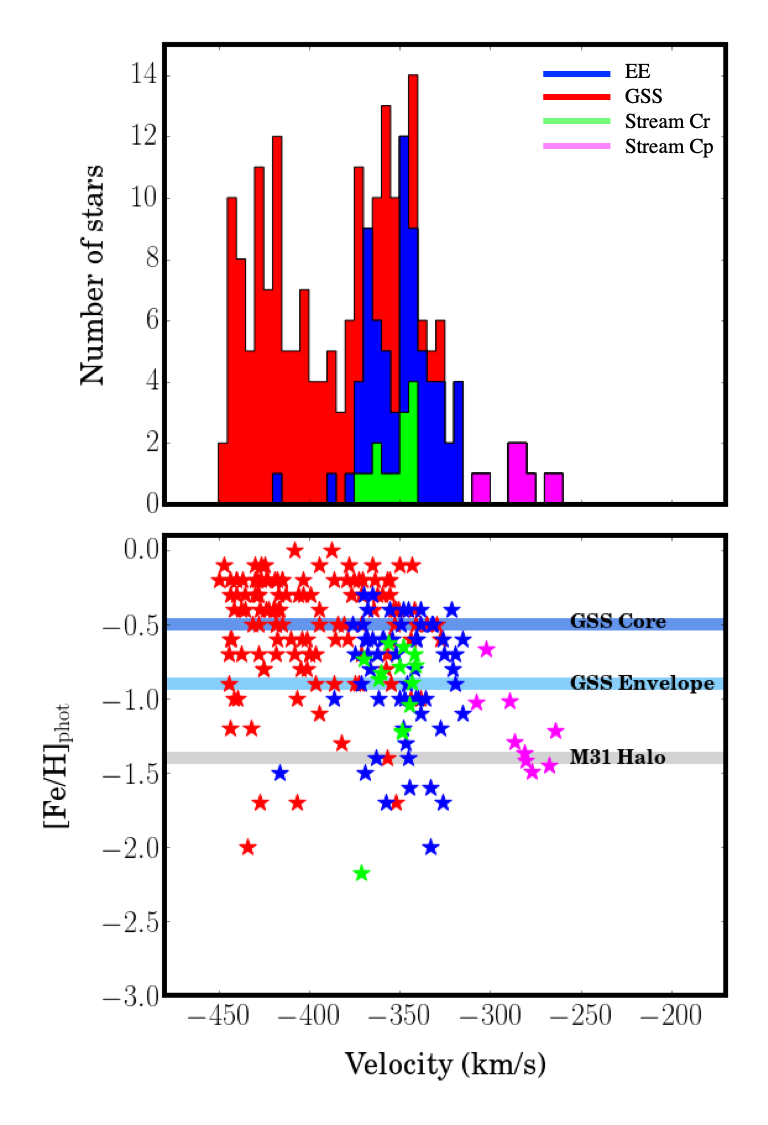}	
    	\vspace*{-10mm}\caption[Metallicity and stellar velocity for Eastern Extent and Giant Stellar Stream stars]
	{Metallicity and stellar velocity for EE and GSS stars. The stacked histogram shows the distribution of stars within the stellar populations for the EE (blue), the GSS(red), Stream Cr (green) and Stream Cp (magenta).  The lower panel plots metallicity vs stellar velocity for the same populations. The horizontal lines indicate previously published values of the $\langle$[Fe/H]$_{\rm phot}$$\rangle$  for key features in the M31 halo. EE/GSS [Fe/H]$_{\rm phot}$ values are derived using isochrones with t = 9 Gyrs, {\alfab}= 0.0 corrected to an heliocentric distance of 845 kpc.}
	\label{EE_Fig21}
\end{figure}

From these results we see that the largest effect on the [Fe/H]$_{\rm phot}$ is $\pm$0.13 dex.  We also note that there is good agreement between the spectroscopic and photometric metallicities for the EE indicating there is no major bias arising from our choice of isochrones. We present our metallicity results in Table \ref{EE_table:20} and 
Figure \ref {EE_Fig21}.  

\begin{table}
	\centering
	\setlength\extrarowheight{2pt}	
	\caption[Mean photometric metallicities for Eastern Extent and Giant Stellar Stream stars ]
	{Mean photometric metallicities for the EE, GSS and Stream C (from \citealt{RefWorks:61}) stellar populations by field. The  $\langle$[Fe/H]$_{\rm phot}$$\rangle$  values are derived using isochrones with \mbox{t = 9 Gyrs}, {\alfab} = 0.0 corrected to an heliocentric distance of 845 kpc. The spectroscopic metallicities for the EE are derived from stacked spectra in each field. }		
	\label{EE_table:20}
	\begin{tabular}{lcc} 
		\hline
		{Field} & $\langle$[Fe/H]$_{\rm phot}$$\rangle$ & $\langle$[Fe/H]$_{\rm spec}$$\rangle$   \\   
		\hline
Af1 &   $-$0.7 $\pm$ 0.3  &  $-$0.9 $\pm$ 0.3   \\ 
Af2 &   $-$1.0 $\pm$ 0.4  &  $-$0.9 $\pm$ 0.3   \\ 
Af3 &   $-$0.9 $\pm$ 0.5  &  $-$0.6 $\pm$ 0.6   \\ 
Af4 &   $-$0.8  $\pm$ 0.5 &  $-$0.9 $\pm$ 0.4   \\ 
Af5 &   $-$1.0 $\pm$ 0.4  &  $-$1.2 $\pm$ 0.3   \\ 
Af6 &   $-$0.9 $\pm$ 0.4  &  $-$0.6 $\pm$ 2.9   \\ 
Af7 &   $-$0.7  $\pm$ 0.3 &  $-$1.3 $\pm$ 0.5   \\ 		
\hline
All EE fields &  $-$0.9 $\pm$ 0.1  & $-$0.9 $\pm$ 0.3   \\  
\hline
S01 &   $-$0.6 $\pm$ 0.3  & \\
S02 &   $-$0.4 $\pm$ 0.3  & \\
S06 &   $-$0.4 $\pm$ 0.1  & \\
S24 &   $-$0.9 $\pm$ 0.5  & \\
S26 &   $-$0.6 $\pm$ 0.3  & \\
S27 &   $-$0.5 $\pm$ 0.4  & \\
\hline
All GSS fields &   $-$0.5 $\pm$ 0.03&   \\
\hline
% Taken from Chapman et al Ref 61
Cr   &   $-$0.7 $\pm$ 0.2 & \\
Cp  &   $-$1.25 $\pm$ 0.2 & \\
		\hline
		\hline
	\end{tabular}
\end{table}

\begin{table}
	\centering
	\setlength\extrarowheight{2pt}	
	\caption[Heliocentric distances for EE and GSS fields ]
	{Heliocentric distances for EE and GSS fields. The data includes field name, the number of the brightest star in the field, its \textit{i}-magnitude, and the distance modulus and heliocentric distance (in kpc) for the field. NB these latter two columns provide an estimate of an upper limit for the distances to the fields by assuming that the brightest star in the confirmed EE/GSS stellar population of each field is at the TRGB.}		
	\label{EE_table:30}
	\begin{tabular}{lcccc} 
		\hline
		Field & Star & \textit {i}-mag & Distance & Distance   \\
		         &        &                       & Modulus & Limit  \\    
		         &         &                      & Limit       & (kpc)   \\   
		\hline
EE fields & & &  & \\
Af1 &  12  &  21.60  &  25.04  &  1019.06  \\  
Af2 &  33  &  21.50  &  24.94  &    974.99  \\   
Af3 &  31  &  21.52  &  24.96  &    979.94  \\  
Af4 &  10  &  21.40  &  24.84  &    930.25  \\   
Af5 &  10  &  21.78  &  25.22  &   1107.64 \\    
Af6 &  25  &  21.35  &  24.79  &    908.66  \\    
Af7 &  24  &  21.35  &  24.79  &    907.82  \\   	
\hline
GSS fields & & &  & \\
S01 &  60 &  21.57  &  25.01  &  1003.23  \\  
S02 &    6 &  21.44  &  24.88  &    947.55  \\  
S06 &    9 &  21.85  &  25.29  &   1143.40  \\    
S24 &  12 &  21.45  &  24.89  &    950.17  \\  
S26 &   69 & 21.35  &  24.79  &    906.98  \\     
S27 &  19 & 21.31  &  24.75  &    889.61  \\     
		\hline
		\hline
	\end{tabular}
\end{table}

As our findings indicate that the EE and GSS exhibit similar kinematics and photometric metallicities, it is possible they may have originated from the same progenitor. If so then we would expect them to have similar trajectories. \cite{RefWorks:148} and \cite{RefWorks:111} found the GSS to lie well behind M31 in the south-eastern quadrant of M31's halo indicative of a trajectory falling in towards M31 from behind. To see if the same is true of the EE we adopt the Tip of the RGB (TRGB) approach to estimating heliocentric distances. The TRGB is a well-accepted standard candle for estimating the heliocentric distances of stellar structures (\citealt{RefWorks:183}, \citealt{RefWorks:45, RefWorks:19, RefWorks:111}, \citealt{RefWorks:334}, \citealt{RefWorks:434}, \citealt{RefWorks:436}, \citealt{RefWorks:535}, \citealt{RefWorks:766}). It relies on the internal processes within a star that cause its luminosity to increase until Helium burning ignites in its core, resulting in a Helium-flash, before the star gradually fades and evolves onto the Horizontal Branch. \cite{M:786} and \cite{M:785}  noted that the onset of Helium burning occurred at an I-band luminosity corresponding to the TRGB and used the bolometric luminosity function to determine the discontinuity indicative of the TRGB.  

For the GSS, \cite{RefWorks:148} and \cite{RefWorks:111} found the TRGB to be i$_0$ $\sim$ 20, which is outside the range we used to select target stars for the observations, so it is unlikely that we will be able to obtain heliocentric distances for our fields that would be consistent with other works. However, for the purposes of comparing distances along the two streams it is possible to obtain an upper limit for the distances to our fields by assuming that the brightest star in the confirmed EE/GSS stellar population of each field is at the TRGB (an approach previously adopted by \citealt{RefWorks:52} and \citealt{RefWorks:175}). We then derive the distances to each field using:

\begin{equation} 
	\label{eq:22a}
	\mathrm{D}{_{\odot}} = 10^{((5 + i_{\mathrm{TRGB}} - M_{\mathrm{TRGB}}) /5) }
\end{equation}	
where: {\Dsun} is the heliocentric distance to the field in parsecs, $i_{\mathrm{TRGB}}$ is the i-magnitude of the brightest star in the EE/GSS stellar population in the field, and $M_{\mathrm{TRGB}}$ is an assumed absolute magnitude for the TRGB of -3.44, which is appropriate for the intermediate to metal-poor stellar populations in our fields (\citealt{RefWorks:111}).  

Our results, presented in Table \ref{EE_table:30}, show that, disregarding the anomalously high value derived for field S06, the distances long the GSS follow a trend consistent with that found by \cite{RefWorks:148} and \cite{RefWorks:111} and consistent with a trajectory for a progenitor approaching M31 from behind.  For the EE we see that the fields all appear to lie behind M31 with {\Dsun} decreasing in the direction from Af6 to Af1 (excepting the anomalous value for Af5), which is consistent with the trajectory for the GSS progenitor.  The distance values derived here are higher than, but of the same order of magnitude as those determined for the GSS by  \cite{RefWorks:148} and for Stream C \cite{RefWorks:111}.  While these results do not conclusively associate the EE, GSS and Stream Cr, they do not rule out the possibility that they were produced by the same progenitor.

%%%%%%%%%%%%%%%%%%%%%%%%%%%%%%%%%%%%%%%%%%%%%%%%%%%%%%%%%%%%%%%%%%%%%%%%%%%
%%%%%%%%%%%%%%%%%%%%%%%%%%%%%%%%%%%%%%%%%%%%%%%%%%%%%%%%%%%%%%%%%%%%%%%%%%%
\section{Discussion} \label{Discussion} 
The results of our kinematic and photometric analysis are shown in the table at Appendix A. It is interesting to note that while the table records no GSS stars present in field S08, we did find six potential candidate stars that had a probability of association with the GSS of $\sim$30\%.  However all of them had a much higher (> 60\%) probability of association with the M31 halo.  This is unsurprising given the on-sky location of this field, which lies well within the M31 halo.  We, therefore, reject these stars as GSS stars and exclude them from further analysis. As a result we are left with no GSS stars in this field.  As it was targeted in the same manner as the others, we assume that there is either a gap in the stream at this location, that the stars are indistinguishable from M31 halo stars or, more likely, that we are overwhelmed by M31 halo stars, hence the low values for P$_{\rm feat}$.

\begin{figure}[h]
  	\centering
	\includegraphics[width=\columnwidth]{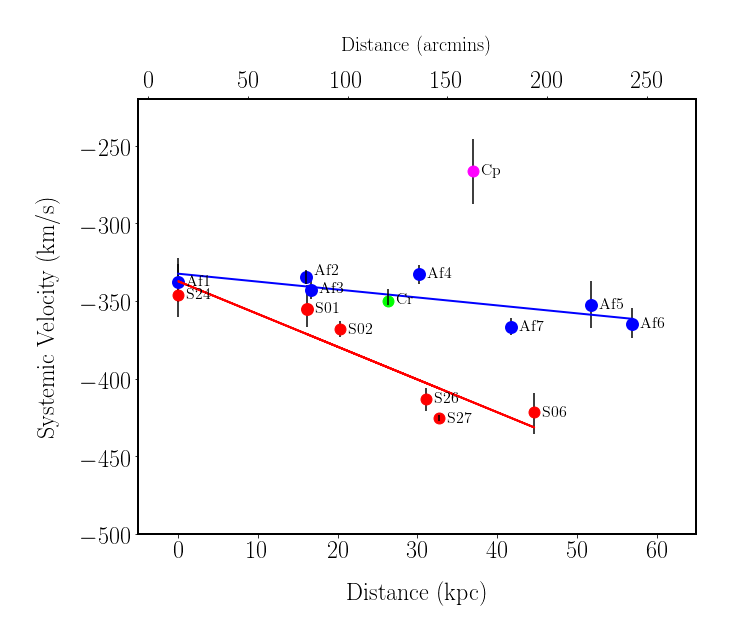}
    	\vspace*{-9mm}\caption[Velocity gradients across Eastern Extent and Giant Stellar Stream fields]
	{Velocity gradients across EE and GSS fields. The velocity gradients are determined with respect to the centre of fields Af1 (for the EE, blue icons) and S24 (for the GSS, red icons) as they lie at one end of each feature and are proximate each other. The values for the velocities and the error bars are obtained from the {\sc emcee} algorithm (\citealt{RefWorks:63}). The field locations are obtained using the mean value of all the $\alpha$s and $\delta$s for the stars in each respective field. The distances are measured from field Af1 for the EE, Stream Cr and Stream Cp and from field S24 for the GSS. The blue line is the best fit line for the EE stars and has a gradient of \mbox{0.51$\pm$0.2 {\kms} kpc$^{-1}$}, in the direction of the GSS, and an intercept $\sim$-332 {\kms}.  The red line is the best fit line for the GSS stars and has a gradient of \mbox{$-$2.11$\pm$0.5 {\kms} kpc$^{-1}$}, in the direction of M31 and an intercept $\sim$-337 {\kms}. The plot also shows the systemic velocities for Stream Cr (green icon) and Stream Cp (magenta icon). Both are obtained by taking the average of the systemic velocities for the respective substructures as recorded by \cite{RefWorks:61}.}
	\label{EE_Fig7}
\end{figure}

%------------------------------------------------------------------------------------------------------------------------------------------------------------
\subsection{Kinematics}\label{Kinematics}
The results, presented in Section \ref{Systemic Velocities}, show the secure EE stellar population has systemic velocities of $-$368 {\kms} $\lesssim$ $\textit v_r$ $\lesssim$$-$331 {\kms}, see Table \ref{EE_table:7}. When we plot these velocities as a function of distance from field Af1 (chosen as an end point of the EE, as on-sky, it is closest to field S24 at the end of the GSS furthest from M31) we see a velocity gradient of $-$0.51$\pm$0.2 {\kms} kpc$^{-1}$ along the EE that is increasing in the direction of field Af1, see Figure \ref{EE_Fig7}.

\begin{figure}
  	\centering
	\includegraphics[width=\columnwidth]{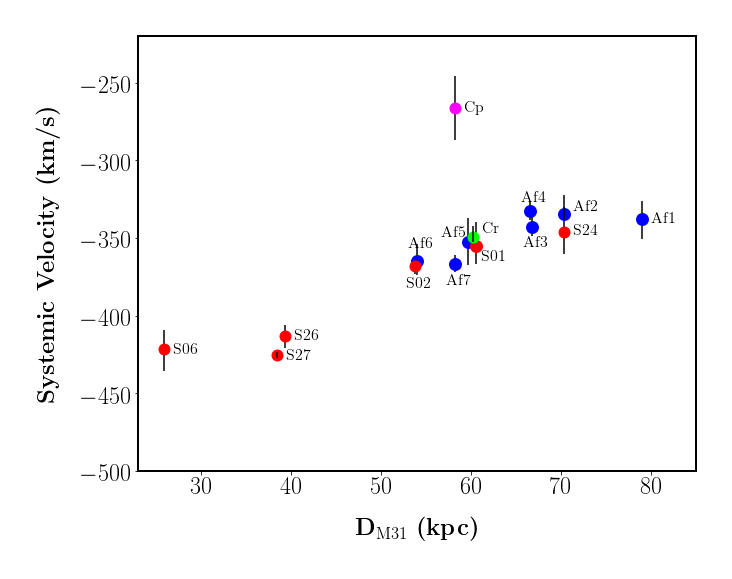}
    	\vspace*{-9mm}\caption[Systemic velocities as a function of radial distance from the centre of M31]
	{Systemic velocities for EE and GSS fields as a function of radial distance from the centre of M31. The values for the velocities and the error bars are obtained from the {\sc emcee} algorithm (\citealt{RefWorks:63}). The field locations are obtained using the mean value of all the $\alpha$s and $\delta$s for the stars in each respective field. EE fields are denoted in blue. GSS fields are shown in red. The plot also shows the systemic velocities for Stream Cr (green icon) and Stream Cp (magenta icon). Both are obtained by taking the average of the systemic velocities for the respective substructures as recorded by \cite{RefWorks:61}.}
	\label{EE_Fig70}
\end{figure}

When we examine our results for the GSS stellar population, we find it has systemic velocities in the range \mbox{$-$431 {\kms}}  $\lesssim$ $\textit v_r$ $\lesssim$$-$346 {\kms}, see Table \ref{EE_table:7}. This is consistent with results from \cite{RefWorks:483},  \cite{RefWorks:38}, \cite{RefWorks:447} and \cite{RefWorks:11}.  In Figure \ref{EE_Fig7}, which plots the systemic velocities of the GSS fields as a function of distance from field S24, we detect a velocity gradient of $-$2.11$\pm$0.5 {\kms} kpc$^{-1}$ along the GSS, increasing in the direction of M31 that is consistent with  \cite{RefWorks:35}.  

In Figure \ref{EE_Fig70} we compare the systemic velocities of the EE and GSS fields and find they are consistent at similar radial distances from the centre of M31. This, and the on-sky proximity of EE field Af1 and GSS field S24 located at the ends of their respective features, suggests the possibility of a physical connection between the two features.

To explore the nature of this potential physical connection we return to Figure \ref{EE_Fig7} and note that with best fit line intercepts of $\sim$$-$332 {\kms} for the EE and $\sim$$-$337 {\kms} for the GSS, fields Af1 and S24 could be within the turning point (see the black open diamond icon on Figure \ref{EE_Fig14}) predicted by \cite{RefWorks:331} from their test particle simulations of the GSS. While this predicted turning point is further along($\sim$20 kpc) the GSS than Af1/S24, the locations are broadly consistent.   If this is the turning point in the stream we would expect to see changes in radial velocity e.g. the slowing down of the velocities along the EE as they yield to the increasing influence of the M31 gravitational potential before turning, gaining speed along the GSS in the direction of M31, which we do.  We would not necessarily expect any significant changes in the metallicities of the stars around the turning point, which is also consistent with our results, i.e. [Fe/H]$_{\rm phot}$ $\sim$ $-$0.7 for field Af1 and [Fe/H]$_{\rm phot}$ $\sim$$-$0.9 for field S24. In terms of the shape of streams at turning points, there are no definitive morphologies. The stream could fan out or could maintain a consistent width.  The determining factor is most likely to be the intrinsic properties of the progenitor, as in the case of NGC 1097, where the internal rotation of the progenitor was a key factor in the stream's abrupt 90$^{\circ}$, \enquote{dog leg}, morphology, \cite{RefWorks:620}.

%------------------------------------------------------------------------------------------------------------------------------------------------------------
\subsection{Photometry}\label{Photometry}
\begin{figure}
  	\centering
	\includegraphics[width=\columnwidth]{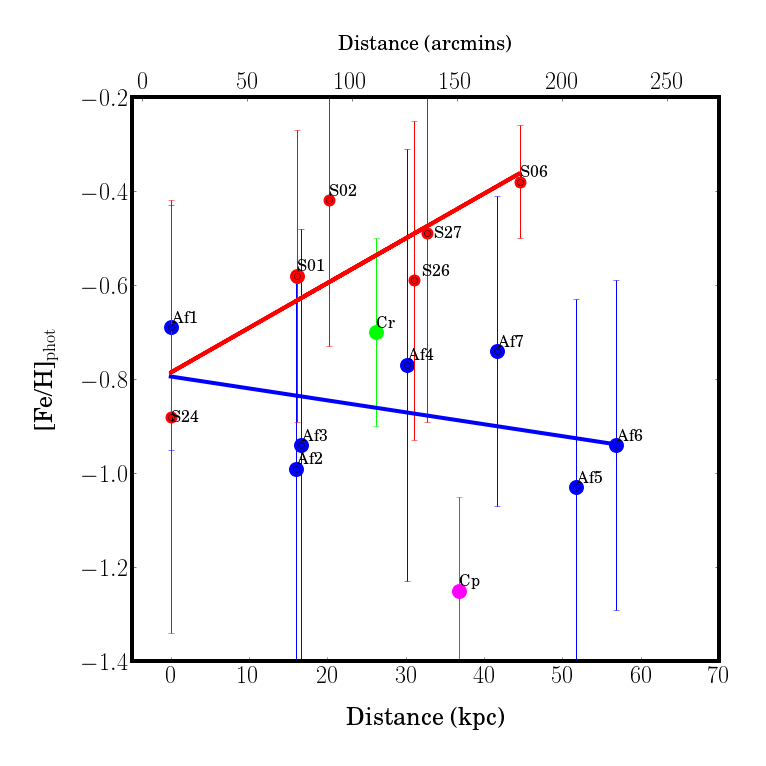}
    	\vspace*{-9mm}\caption[Metallicity gradients across Eastern Extent and Giant Stellar Stream fields]
	{Metallicity gradients for EE and GSS stars. The distances are measured from field Af1 for the EE (blue icons), Stream Cr (green icon) and Stream Cp (magenta icon) and from field S24 for the GSS (red icons). The blue line is the best fit line for the EE fields and has no discernible gradient.  The red line is the best fit line for the GSS fields and has a very small gradient of 0.01$\pm$0.0.005 dex kpc$^{-1}$ in the direction of M31. The  $<$[Fe/H]$_{\rm phot}$$>$ for the Stream C fields  are obtained by taking the average of the metallicities for the respective substructures as recorded by \cite{RefWorks:61}.}
	\label{EE_Fig15}
\end{figure}

Our results for the EE show $-$1.0 $\lesssim$ [Fe/H]$_{\rm phot}$  $\lesssim$ $-$0.7 with an overall mean of $\langle$[Fe/H]$_{\rm phot}$$\rangle$ = $-$0.9 $\pm$ 0.1, see Table \ref{EE_table:20}. We find that this changes little along the length of the feature with Figure \ref{EE_Fig15} showing no discernible metallicity gradient across the fields.

Our results for the GSS show $-$0.9 $\lesssim$ [Fe/H]$_{\rm phot}$ $\lesssim$ $-$0.4 with an overall mean = $-$0.5 $\pm$ 0.4, see Table \ref{EE_table:20}. However, in this instance we find a very small, $-$0.01$\pm$0.005 dex kpc$^{-1}$, metallicity gradient (see Figure \ref{EE_Fig15}) with stars becoming increasingly metal poor with distance from M31.  These results are consistent with  \cite{RefWorks:35}, who found a small gradient of $-$0.0101 $\pm$ 0.005 dex kpc$^{-1}$ and with findings from \cite{RefWorks:483}, \cite{RefWorks:331},  \cite{RefWorks:550}, \cite{RefWorks:447}, \cite{RefWorks:203}, \cite{RefWorks:111} and \cite{RefWorks:495}. 

\begin{figure}
  	\centering
	\includegraphics[width=\columnwidth]{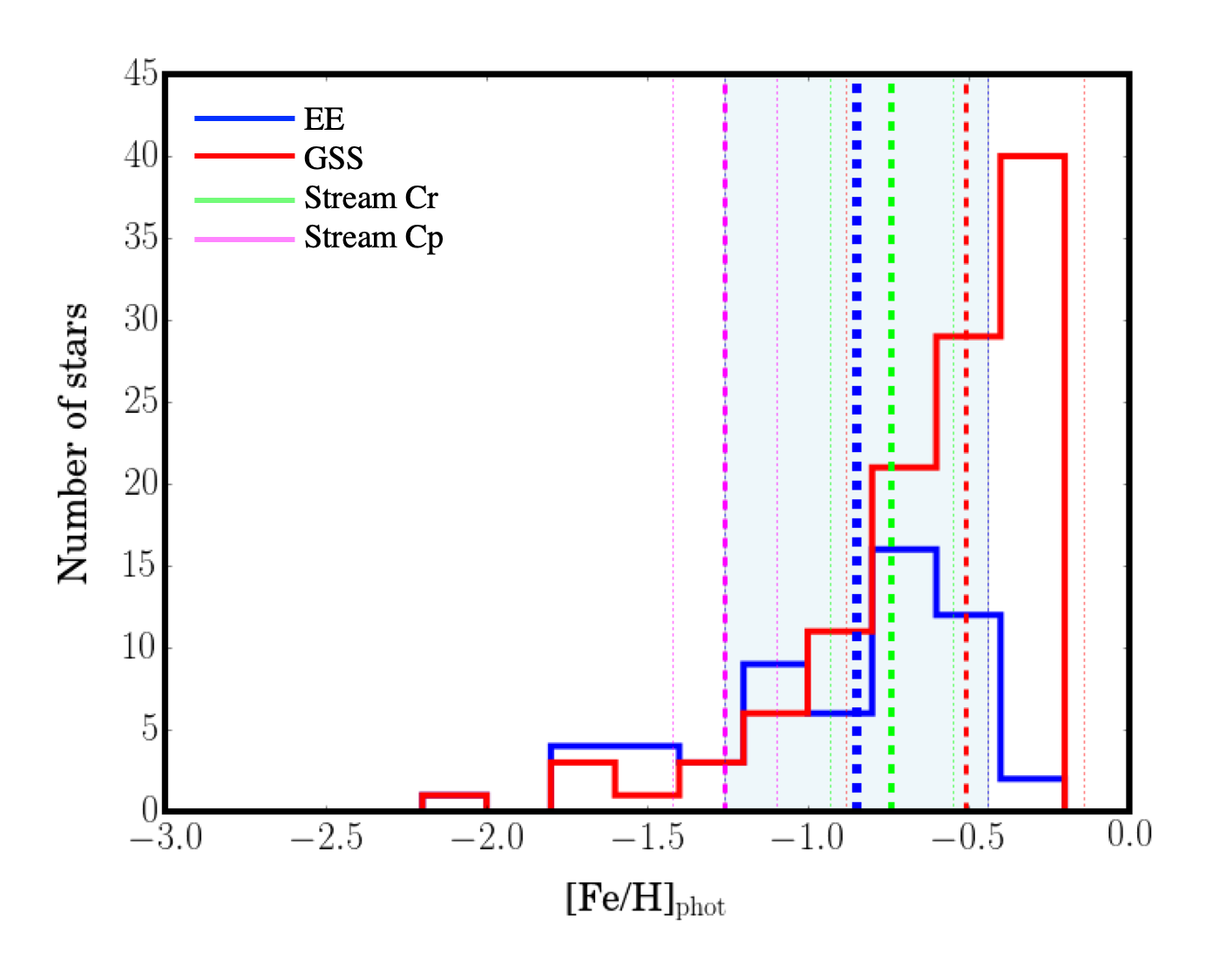}
    	\vspace*{-9mm}\caption[Metallicity distributions for Eastern Extent, Giant Stellar Stream and Stream C stars]
	{Photometric metallicity distributions for EE, GSS  and Stream C stars. The blue outline indicates the [Fe/H]$_{\rm phot}$ distribution of the EE stars and the red outline shows that of the GSS stars. The vertical, dashed, lines show the position of the $\langle$[Fe/H]$_{\rm phot}$$\rangle$ for the EE (blue), the GSS (red) Stream Cr (green) and Stream Cp (magenta). The shaded area shows the extent of the standard deviation for the EE metallicity distribution.  The vertical, dotted, lines indicate the standard deviations for the GSS (red), Stream Cr (green) and Stream Cp (magenta).  The  [Fe/H]$_{\rm phot}$ values are derived using isochrones with t = 9 Gyrs, {\alfab} = 0.0, corrected to an heliocentric distance of 845 kpc.}
	\label{EE_Fig17}
\end{figure}

In Figure \ref{EE_Fig17} we plot the metallicity distributions of the EE and GSS and see that they have similar profiles, i.e. each having a dominant metal rich peak and tails of increasingly metal poor stars, and that their metallicities are consistent to within 1-$\sigma$.  We also note that the EE lacks the metal-rich population that dominates the GSS.

%------------------------------------------------------------------------------------------------------------------------------------------------------------
\subsection{Streams B, C and D}\label{Stream C}
When we look at the positions of other streams in the M31 halo, Stream C is the nearest to the EE (see Figure \ref{EE_Fig14}).  To determine if there is any association between these two features we compare their kinematic and photometric properties.  We use work by \cite{RefWorks:61}, \cite{RefWorks:11}, \cite{RefWorks:203} who find that Stream C comprises two substructures:
 
 \begin{itemize}
	\item Stream Cr: with a v$_r$ = $-$349.5 $\pm$ 1.8 {\kms}, \mbox{$\sigma_v$ = 5.1 $\pm$ 2.5 {\kms}} and a <[Fe/H]> = $-$0.7 $\pm$ 0.2.
	\item Stream Cp: with a v$_r$ = $-$285.6 $\pm$ 1.2 {\kms}, \mbox{$\sigma_v$ = 4.3$^{+1.7}_{-1.4}$ {\kms}} and a <[Fe/H]> = $-$1.3$\pm$ 0.2 - indicating it to be more metal poor than Stream Cr.
\end{itemize}

In addition there is a globular cluster, EC4, co-located on the sky with Stream Cp.  \cite{RefWorks:55} found sufficient similarities in their properties to suggest that EC4 and Stream Cp are related to one another, but that EC4 is unlikely to be this stream's progenitor. When we compare the properties of Streams Cr, Cp, EC4 with those of the EE we find that Stream Cr is kinematically consistent with the EE (see Figure \ref{EE_Fig7}), while Stream Cp and EC4 are not. Similarly, when reviewing Figures \ref{EE_Fig15} and \ref{EE_Fig17}, we see that Stream Cr is similar to the EE, with a $\langle$[Fe/H]$_{\rm phot}$$\rangle$ within the standard deviation of that for the EE. However, Stream Cp has a distinctly different metallicity indicating that it, and by association EC4, are quite different features from the EE.  This means it is plausible that Stream Cr and the EE are related or are actually part of the same debris structure. 

Findings by \cite{RefWorks:111} for heliocentric distances along Stream C are consistent with those of the GSS and with our findings for the EE.  The similarity in the kinematics, photometric metallicities and trajectories of these features are plausible indications that they were formed from a progenitor falling in from behind M31.

With respect to Stream B, its metallicity distribution (see Figure 33 by \citealt{RefWorks:38}) shows it to be a more metal-poor feature than the EE, peaking at [Fe/H] $\sim$$-$1.0 and with a long tail extending out to [Fe/H] $\sim$$-$3.0.  However, when we compare Stream B's systemic velocity, \mbox{$\sim$$-$330 {\kms}}(\citealt{RefWorks:61}) with that of the EE we find them to be consistent, so it is not impossible for these two features to be related.  

Stream D has a similar, metal-poor, profile to Stream B, but with a systemic velocity $\sim$$-$390 {\kms} that is much higher than that of the EE, we conclude it is unlikely that these two features are associated.  

%------------------------------------------------------------------------------------------------------------------------------------------------------------
\subsection{Globular Clusters}\label{Globular Clusters}
Three globular clusters lie within the EE footprint and one lies close to its tip (see LC14 on Figure \ref{EE_Fig14}). \cite{RefWorks:589} find this latter cluster, LAMOST-C14 to have a radial velocity, v$_r$ = 61 {\kms} and [Fe/H] = $-$1.3.  Both of these properties are distinctly different from those exhibited by the EE, indicating that it is unlikely that there is any association between these two features. The remaining three globular clusters are H26 and two halo-extended clusters HEC12 (aka EC4), discussed in Section \ref{Stream C}, and HEC-13, all of which were first reported by \cite{RefWorks:95}.  

With a radial velocity, v$_r$= $-$411 $\pm$ 7 {\kms} (\citealt{RefWorks:95}, \citealt{RefWorks:516}) and metallicity, [Fe/H] = $-$1.6 (\citealt{RefWorks:590}), H26 also appears to be different in nature from the EE. On-sky it is not close to any of the EE fields. The nearest field, Af2, has a systemic velocity of $\sim$$-334$ $\pm$ 3 {\kms}  and a [Fe/H] =$\sim$$-$0.9 which are not consistent with those of H26.  So we discount an association between these two features. We also note that it is unlikely that there is any association between H26 and Stream C as their properties are also inconsistent. This is counter to the view of  \cite{RefWorks:590}.

HEC-13 has a radial velocity, v$_r$ = $-$366 $\pm$ 5 {\kms}, \cite{RefWorks:516} which is consistent within 90\% confidence limits to the systemic velocities of the nearest EE fields, Af2 and Af3, so there is a potential kinematic consistency between the EE and HEC-13. Similarly, this cluster has the potential to be associated with Stream Cr, which has a radial velocity, v$_r$= $-$349.5 $\pm$ 1.8 {\kms}, but it is unlikely to be associated with Stream Cp.

There are also three globular clusters projected onto the GSS: H19, H22 and PAndAS-37.   Our findings indicate that the radial velocity of PAndAS-37 (v$_r$ = -404 $\pm$ 15 {\kms}) is consistent with those measured for nearby fields in the GSS but that the velocities of H19 (v$_r$ = -272 $\pm$ 18 {\kms}) and H22 (v$_r$ = -311 $\pm$ 6 {\kms}) are not.  These findings are in agreement with conclusions drawn by \cite{RefWorks:516}.

%------------------------------------------------------------------------------------------------------------------------------------------------------------
\subsection{The Nature of the EE}\label{The Nature of the EE}
We have shown that the EE overlaps Stream C on the sky and exhibits similar kinematics, photometric metallicities and morphologies to the substructure Stream Cr, so it is possible that these two features are related or comprise the same structure. But what is the nature of that structure - is it a stream or a shell?  

Our findings do not conclusively support either option yet plausibly support both. Results from our metallicity analysis indicate a shallow gradient along the GSS and, if we collate data for the EE and GSS from this work and with that for Streams A, B, C and D from \citealt{RefWorks:61} i.e: 

\begin{tabular}{lll}
Stream A  &  $\langle$[Fe/H]$_{\rm phot}$$\rangle$ = $-$1.3 $\pm$ 0.3 \\
Stream B  &  $\langle$[Fe/H]$_{\rm phot}$$\rangle$ = $-$0.8 $\pm$ 0.2 \\
EE            &  $\langle$[Fe/H]$_{\rm phot}$$\rangle$ = $-$0.9 $\pm$ 0.1 \\
Stream C  &  $\langle$[Fe/H]$_{\rm phot}$$\rangle$ = $-$1.0 $\pm$ 0.2 \\
Stream D  &  $\langle$[Fe/H]$_{\rm phot}$$\rangle$ = $-$1.1 $\pm$ 0.3 \\
GSS         &  $\langle$[Fe/H]$_{\rm phot}$$\rangle$ = $-$0.5 $\pm$ 0.4 \\
\end{tabular}
\\

\noindent we see they are indicative of a shallow metallicity gradient across a substantive section of the M31 halo, consistent with findings by \cite{RefWorks:82} and \cite{RefWorks:35}. The metallicities are also more metal-rich than M31's smooth halo ( <[Fe/H]> $\sim$ -1.7) which led \cite{RefWorks:82} to conclude that the GSS is a recently formed feature.  With the estimated ages of the GSS $\sim$ 9 Gyrs (\citealt{RefWorks:552}) and Stream C $\sim$9.3$^{+0.9}_{-1.4}$ Gyrs (\citealt{RefWorks:203}) and assuming the EE to be a similar age, the relative youth of these features and the shallow metallicity gradient across them would seem to indicate that they are plausibly part of a shell system formed during a major merger of their progenitor and M31.

On the other hand, evidence from our velocity dispersion analysis is consistent with that from \cite{RefWorks:176} who concluded that the progenitor of the GSS was a low mass dwarf galaxy, M = $\sim$ 10$^9$ {\Msun}, that could also be the progenitor of the EE and the other adjacent structures. This could imply that these features are all part of the same debris structure.

This apparent dichotomy can be resolved by the hypothesis that the EE and Streams C and D were the result of the merger of a satellite of the progenitor \mbox{(M = $\sim$10$^{11}$ {\Msun})} of the GSS or were formed during one or more subsequent minor events (\citealt{RefWorks:154, RefWorks:413}, \citealt{RefWorks:470, RefWorks:637}).

While these insights into the properties of the EE, GSS and adjacent streams do not fully resolve the conundrum of how they were formed they do provide many more detailed constraints on their kinematics and composition. These new data could inform future research and deliver greater clarity around the events that brought these exquisite structures around M31 into being.

%%%%%%%%%%%%%%%%%%%%%%%%%%%%%%%%%%%%%%%%%%%%%%%%%%%%%%%%%%%%%%%%%%%%%%%%%%%
%%%%%%%%%%%%%%%%%%%%%%%%%%%%%%%%%%%%%%%%%%%%%%%%%%%%%%%%%%%%%%%%%%%%%%%%%%%
%------------------------------------------------------------------- Conclusions ----------------------------------------------------------------------------------------------------------------
\section{Conclusions} \label{Conclusions} 
In this work we present the first comprehensive spectroscopic survey of M31's Eastern Extent.  We determine the kinematics and photometry for $\sim$50 RGB stars in 7 fields across the EE extending for $\sim$4$^{\circ}$ on-sky at radial distances 70 kpc $\le$ R < 90 kpc from the centre of M31.  We also present a comparison of the properties of these stars with those in Stream C and with $\sim$100 RGB stars in the GSS to determine whether or not there is an association between these features. Here we summarise our key findings:

 \begin{itemize}
	\item The systemic velocities of fields in the EE lie in the range $-$368 {\kms} $\lesssim$ $\textit v_r$ $\lesssim$$-$331 {\kms}, with a slight velocity gradient of $-$0.51$\pm$0.2 {\kms} kpc$^{-1}$ across them.
	\item Metallicities along the EE lie in the range $-$1.0 $\lesssim$ [Fe/H]$_{\rm phot}$ $\lesssim$ $-$0.7 with a $\langle$[Fe/H]$_{\rm phot}$$\rangle$ = $-$0.9 $\pm$ 0.1, with no discernible gradient across the fields.
	\item When we compare the results of the EE with neighbouring Streams Cr and Cp, we find strong similarities between the properties of the EE and those of Stream Cr, plausibly linking the two structures or even indicating that they could both belong to the same feature. However, we find that Stream Cp, and its associated globular cluster, EC4, to have distinctly different properties, indicative of a separate structure. 
	\item Similar comparisons with Streams B and D find there is a tentative association with Stream B and the EE, but not for Stream D and the EE. 
	\item We find a kinematic consistency between the EE and globular cluster HEC-13, however, without additional, corroborating information we have insufficient data to support an hypothesis that these two features are related.
	\item When we compare our results to similar properties of the GSS we find them to be consistent such that the EE could plausibly comprise stars stripped from the progenitor of the GSS. 
\end{itemize}

%%%%%%%%%%%%%%%%%%%%%%%%%%%%%%%%%%%%%%%%%%%%%%%%%%%%%%%%%%%%%%%%%%%%%%%%%%%
%%%%%%%%%%%%%%%%%%%%%%%%%%%%%%%%%%%%%%%%%%%%%%%%%%%%%%%%%%%%%%%%%%%%%%%%%%%
%------------------------------------------------------------------- Acknowledgements -------------------------------------------------------------------------------------------------------
\section{Acknowledgements}
The authors wish to thank the anonymous referee for their helpful comments and suggestions.  JP wishes to thank B.Sullivan, J. Sullivan and S. Sullivan for their support. Most of the data presented herein were obtained at the W.M. Keck Observatory, which is operated as a scientific partnership among the California Institute of Technology, the University of California and the National Aeronautics and Space Administration. The Observatory was made possible by the generous financial support of the W.M. Keck Foundation. The authors wish to recognise and acknowledge the very significant cultural role and reverence that the summit of Mauna Kea has always had within the indigenous Hawaiian community. 

%%%%%%%%%%%%%%%%%%%%%%%%%%%%%%%%%%%%%%%%%%%%%%%%%%%%%%%%%%%%%%%%%%%%%%%%%%%
%%%%%%%%%%%%%%%%%%%%%%%%%%%%%%%%%%%%%%%%%%%%%%%%%%%%%%%%%%%%%%%%%%%%%%%%%%%
%------------------------------------------------------------------- Data Availability ------------------------------------------------------------------------------------
\section{Data Availability}
The data underlying this article are available in the article and in its online supplementary material. The raw DEIMOS data are available via the Keck archive. The reduced spectra are available on request from the lead author.\\

%%%%%%%%%%%%%%%%%%%%%%%%%%%%%%%%%%%%%%%%%%%%%%%%%%%%%%%%%%%%%%%%%%%%%%
%%%%%%%%%%%%%%%%%%%%%%%%%%%%%%%%%%%%%%%%%%%%%%%%%%%%%%%%%%%%%%%%%%%%%%	
%------------------------------------------------------------------- Bibliography  ------------------------------------------------------------------------------------
\bibliography{BiblogEE}
\bibliographystyle{mnras}

\end{document}